\documentclass[journal]{vgtc}                

\ifpdf
  \pdfoutput=1\relax                   
  \usepackage{graphicx}                
  \DeclareGraphicsExtensions{.pdf,.png,.jpg,.jpeg} 
\else
  \usepackage{graphicx}                
  \DeclareGraphicsExtensions{.eps}     
\fi%

\graphicspath{{figures/}{pictures/}{images/}{fig/}{./}} 

\usepackage{comment}
\usepackage{amsmath} 
\usepackage{microtype}                 
\usepackage{dblfloatfix} 
\PassOptionsToPackage{warn}{textcomp}  
\usepackage{textcomp}                  
\usepackage{mathptmx}                  
\usepackage{times}                     
\usepackage{cite}                      
\usepackage{tabu}                      
\usepackage{booktabs}                  
\usepackage{siunitx}

\onlineid{1129}

\vgtccategory{Research}
\vgtcpapertype{Evaluation}

\usepackage[dvipsnames]{xcolor}
\usepackage{booktabs, tabularx, colortbl} 
\usepackage{wrapfig, multirow, amssymb}
\usepackage{xcolor}

\newcommand{\bbc}[1]{\textcolor{RoyalBlue}{(Bobby: #1})}
\newcommand{\haley}[1]{\textcolor{OrangeRed}{(Haley: #1)}}

\newcommand{\jeanine}[1]{\textcolor{Violet}{(Jeanine: #1)}}

\newcommand{\sarah}[1]{\textcolor{OliveGreen}{(Sarah: #1)}}

\title{Stay in Touch! Shape and Shadow Influence Surface Contact in \\ XR Displays}

\author{Haley Adams, Holly Gagnon, Sarah Creem-Regehr, Jeanine Stefanucci, and Bobby Bodenheimer, \textit{Member, IEEE}} 
\authorfooter{
\item
 Haley Adams is with the Department of Computer Science Vanderbilt University. E-mail: haley.a.adams@vanderbilt.edu.
\item
 Holly Gagnon is with the Department of Psychology at University of Utah. E-mail: holly.gagnon@psych.utah.edu.
\item
 Sarah Creem-Regehr is with the Department of Psychology at University of Utah. E-mail: sarah.creem@psych.utah.edu.
\item
 Jeanine Stefanucci is with the Department of Psychology at University of Utah 
 Research. E-mail: jeanine.stefanucci@psych.utah.edu.
 \item
 Bobby Bodenheimer is with the Department of Computer Science Vanderbilt University. E-mail: bobby.bodenheimer@vanderbilt.edu.
}

\shortauthortitle{Adams \MakeLowercase{\textit{et al.}}: Shape \& Shadow in XR}

\abstract{The information provided to a person's visual system by extended reality (XR) displays is not a veridical match to the information provided by the real world. Due in part to graphical limitations in XR head-mounted displays (HMDs), which vary by device, our perception of space may be altered. However, we do not yet know which properties of virtual objects rendered by HMDs---particularly augmented reality displays---influence our ability to understand space. In the current research, we evaluate how immersive graphics affect spatial perception across three unique XR displays: virtual reality (VR), video see-through augmented reality (VST AR), and optical see-through augmented reality (OST AR). We manipulated the geometry of the presented objects as well as the shading techniques for objects' cast shadows. Shape and shadow were selected for evaluation as they play an important role in determining where an object is in space by providing points of contact between an object and its environment---be it real or virtual. 
Our results suggest that a non-photorealistic (NPR) shading technique, in this case for cast shadows, may be used to improve depth perception by enhancing perceived surface contact in XR. Further, the benefit of NPR graphics is more pronounced in AR than in VR displays. One's perception of ground contact is influenced by an object's shape, as well. However, the relationship between shape and surface contact perception is more complicated.  
} 


\CCScatlist{ 
  \CCScat{I.3.7}{Computer Graphics}%
{Three-Dimensional Graphics and Realism}{Virtual Reality};
  \CCScat{J.4}{Computer Applications}{Social and Behavioral Sciences}{Psychology}
}

\teaser{
  \centering
  \includegraphics[width=\linewidth]{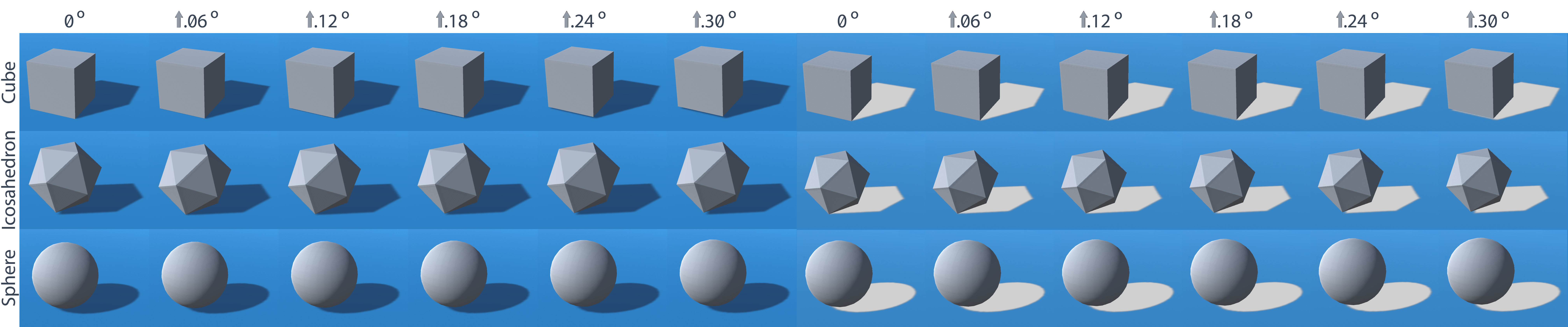}
\caption{Three geometric shapes were evaluated in Experiment 1. Each object was rendered with either dark (photorealistic) or light (non-photorealistic) shadow shading methods and was displaced vertically based on the viewer's visual angle to the target. Changes in vertical displacement were subtle since a psychophysical, two-alternative forced choice (2AFC) approach was employed to evaluate surface contact perception.  }
	\label{fig:exp1_all}
}

\vgtcinsertpkg

\begin{document}


\firstsection{Introduction}

\maketitle

\maketitle


Rendering methods for extended reality (XR) systems are improving, and many displays can now render compelling virtual objects. Some difficulties still remain, however, and one that is common in consumer-level augmented reality (AR) is that objects may seem detached from their surroundings. For example, a Pok\'{e}mon in Pok\'{e}mon Go, the popular mobile AR game, may seem to float above the ground where it is placed. A virtual hat in an AR face filter, like those used by Snapchat or Instagram, may appear detached from a person’s head. This phenomenon is a strong indicator that the depth information presented by virtual stimuli and the information presented by the real world environment do not sufficiently match. In the current research, we evaluate how the way we render cast shadows for objects of different geometric shapes and orientations affects one’s perception of surface contact across different extended reality (XR) displays.

Recent work on optical see-through (OST) AR displays leverages the human visual system to create shadows~\cite{Manabe:2019:SII, Ikeda:2020:SIO}. These devices create the illusion of dark color values using simultaneous contrast illusion, which approximates a shadow. Other methods have been used for video see-through (VST) AR displays~\cite{Noh:2009:RST}, and shadow generation in virtual reality (VR) is considered mature~\cite{Slater:1995:IDS,Eisemann:2009:CSR}. Nonetheless, for generally deployable XR applications, it would be desirable to have a clear and unified understanding of which techniques work across widely available technologies. Therefore, in the present work we investigate how surface contact is affected by cast shadows across a set of display types.

Additionally, Steptoe et al.~\cite{Steptoe:2014:PDC} recently made a compelling case for non-photorealistic shadow applications in AR.  Some prior work suggests that non-photorealistic shadow shading techniques in AR may be beneficial for attaching virtual objects to the surfaces beneath them \cite{Adams:2021:SLC}. Thus, we also investigate how dark, photorealistic and light, non-photorealistic cast shadows affect surface contact judgments in XR. The possibility of non-photorealistic shadows offering benefits for surface contact judgments may provide additional flexibility for XR application designers given the difficulty in displaying dark shadows that exists in many AR devices.


In the current work, we make several contributions by evaluating how people's judgments are affected by virtual object shape, orientation, and shadow shading in three different XR displays. We demonstrate that non-photorealistic, light shadow shading can enhance surface contact judgments. We show that these judgments are complex and affected by an object's shape as well as the orientation in which it is presented. Our results also lay groundwork to encourage further study of non-photorealistic rendering techniques to improve spatial perception in XR.

\begin{figure}[t]
    \centering
        \includegraphics[width=0.49\linewidth, keepaspectratio]{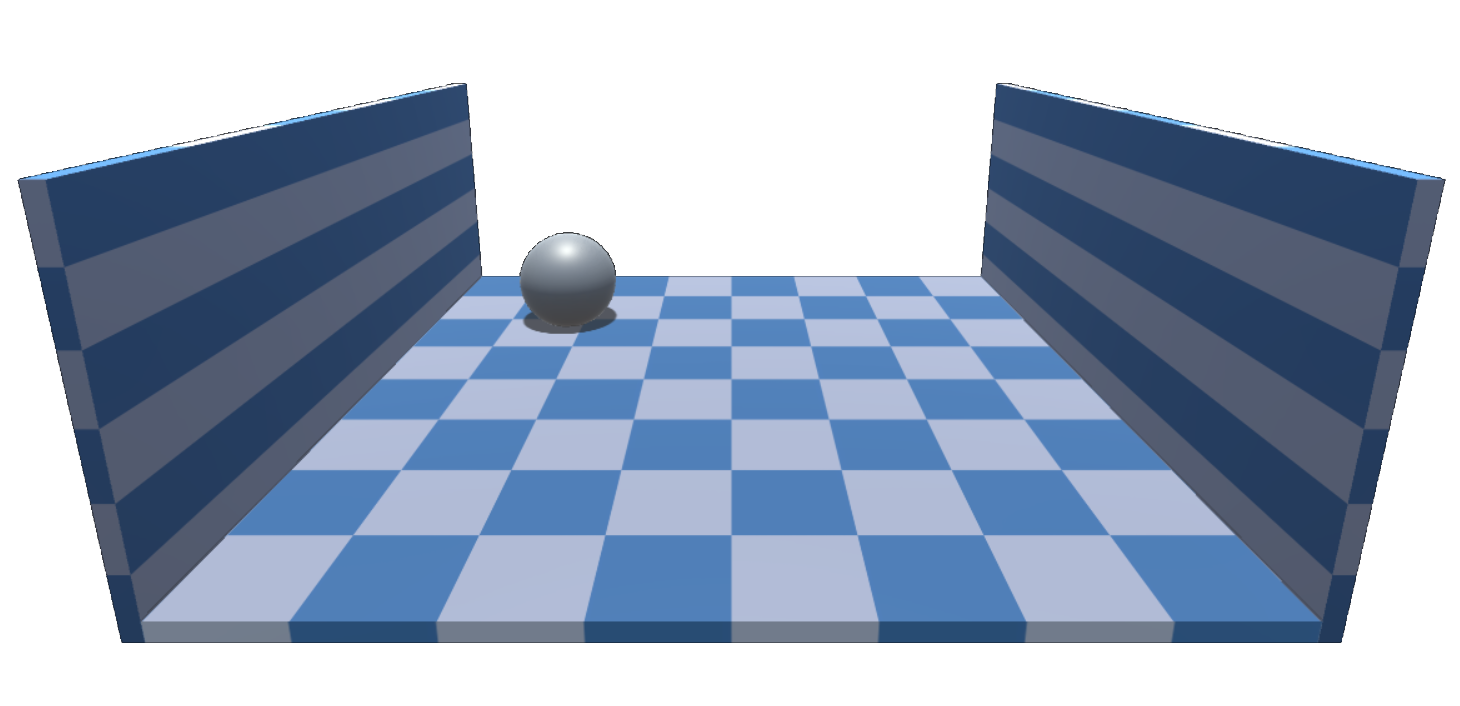}
        \includegraphics[width=0.49\linewidth, keepaspectratio]{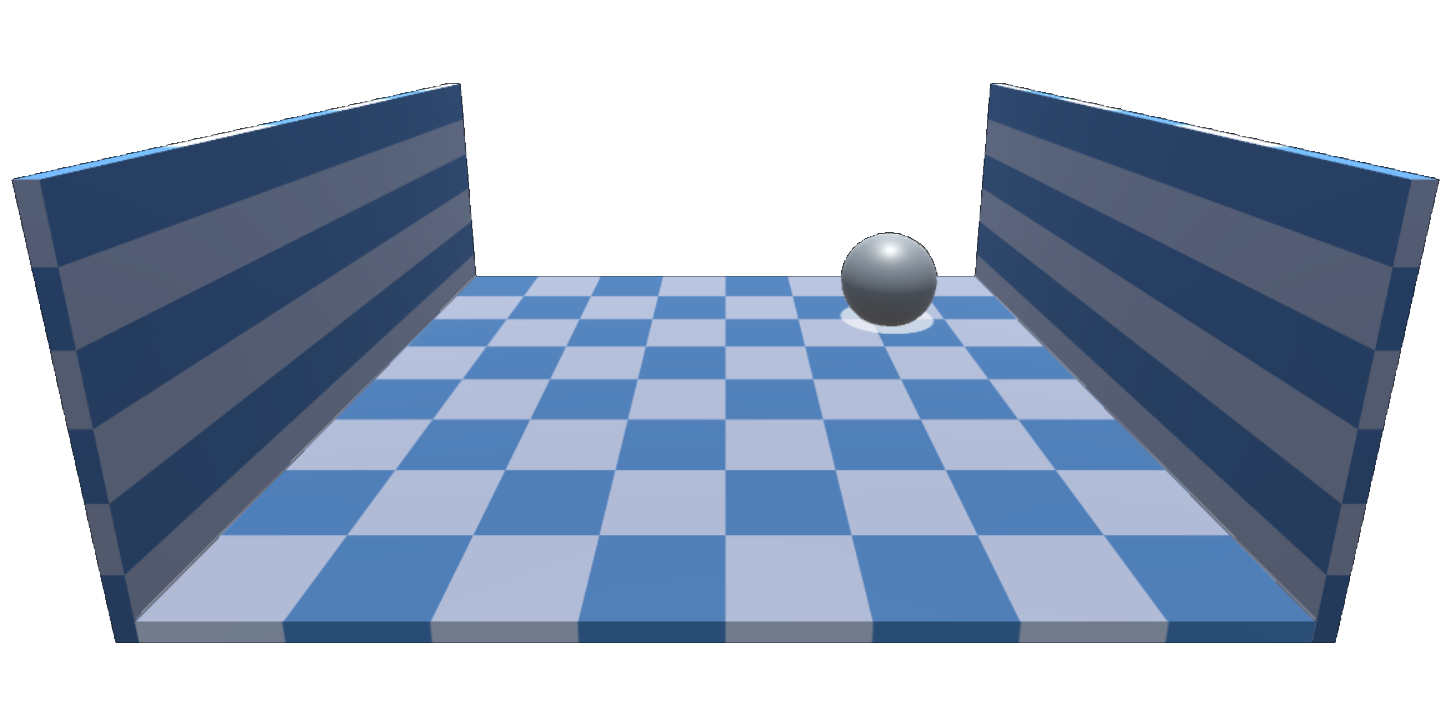}
        \includegraphics[width=0.49\linewidth, keepaspectratio]{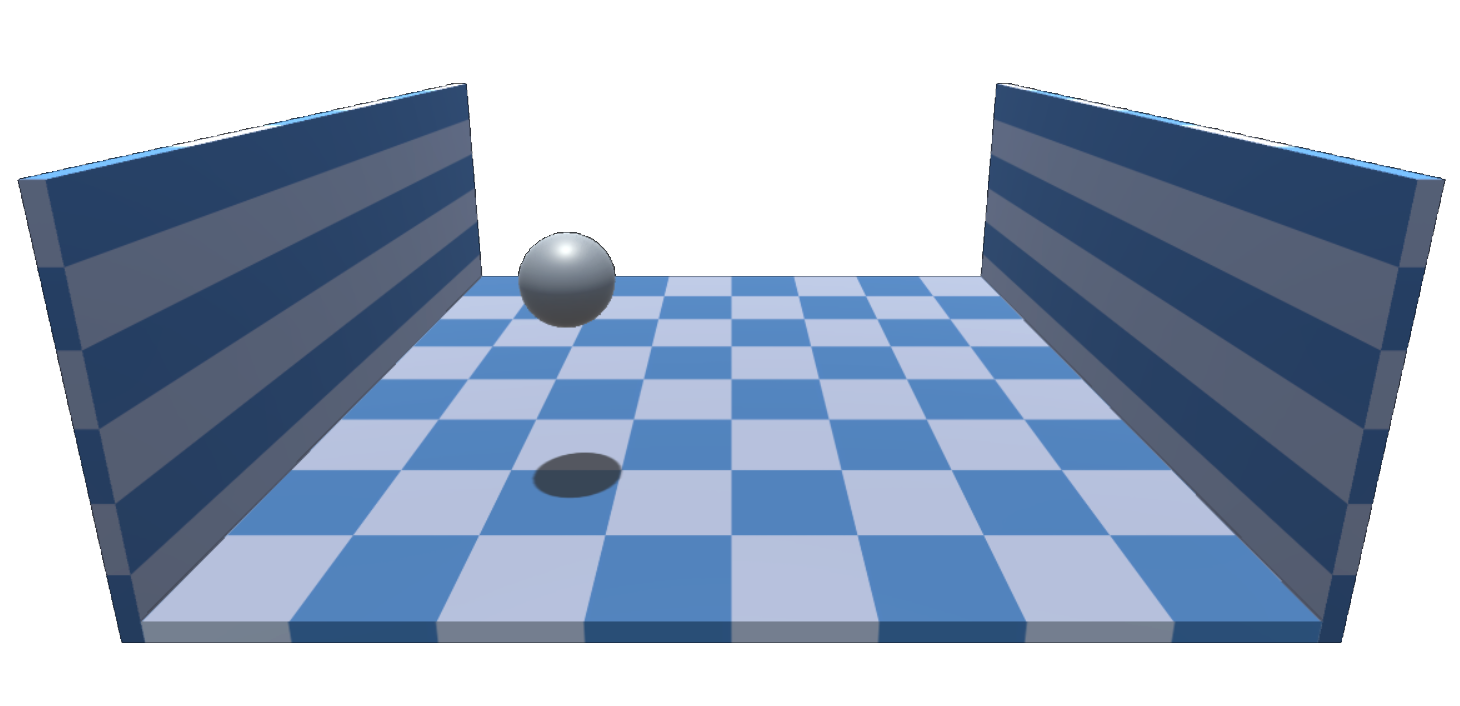}
        \includegraphics[width=0.49\linewidth, keepaspectratio]{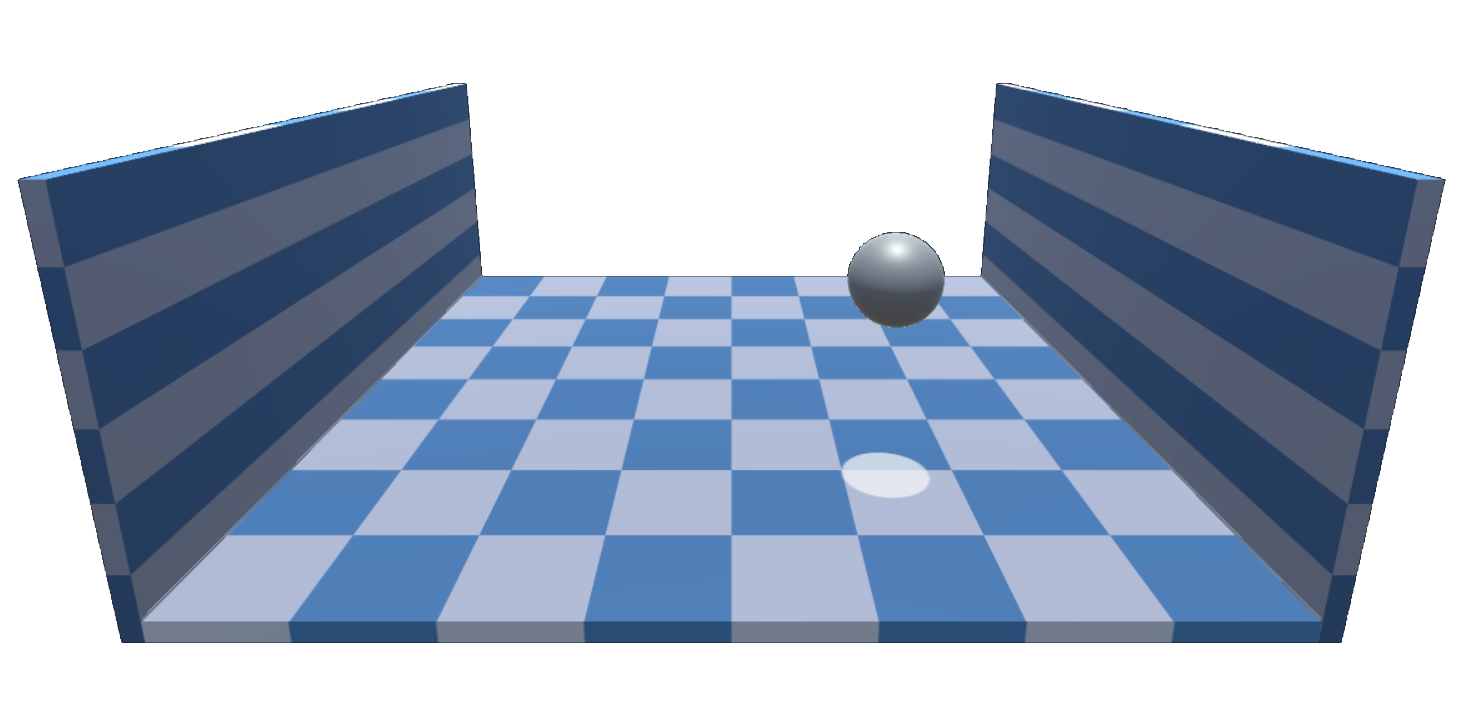}
        \caption{A recreation of the ``ball-in-a-box'' demonstration \cite{Kersten:1997:MCS}. Although the sphere is located at the same position in all four images, it appears closer in the bottom images due to the placement its cast shadow. The influence of cast shadow on an object's position works regardless of whether the shadow is rendered realistic (left) or nonrealistic (right) manner. Top:~the sphere seems to rest along the ground due to an anchoring shadow beneath it that connects the sphere with the back of the box. Bottom:~the sphere seems to float above the ground with a detached shadow beneath it that connects it towards the front of the box. }
        \label{fig:ball-in-box}
\end{figure}

\section{Related Work}

\subsection{Cast Shadows} 

Cast shadows---from which humans infer size \cite{Yonas:1978:DSI,Wanger:1992:ESQ, Casati:2016:SFS}, shape \cite{Cavanagh:1989:SFS, Knill:1997:GS}, and the distance between an object and adjacent surfaces\cite{Thompson:1998:VG, Hu:2000:VCI, Madison:2001:UIS}---play a particularly important role in our perception of space. In fact, one's perception of the position of an object can shift dramatically depending on the placement of its associated cast shadow \cite{Mamassian:1998:PCS, Ni:2004:PSL}.  Kersten et al.'s ``ball-in-a-box'' study provides an important example of this effect \cite{Kersten:1997:MCS}. Kersten and colleagues demonstrated that by changing the location of a shadow in space, even a stationary object may appear to move (See Figure \ref{fig:ball-in-box}). Furthermore, their research demonstrated that cast shadow shading could be manipulated with light, photometrically incorrect shading, but still provide a cue to determine spatial location. For a recent and thorough review of shadow perception, see Casati and Cavanagh \cite{Casati:2019:VWS}.






\subsection{Shape \& Shadows in XR } 


In traditional computer graphics, evidence shows that cast shadows function as ``visual glue'' to attach virtual objects to surfaces \cite{Thompson:1998:VG,Madison:2001:UIS}.  However, it is unclear how to best create visual glue for augmented reality devices. AR introduces new complexities to the graphics pipeline, especially for OST displays, which rely on additive light for rendering. These displays cannot remove light, and thereby darken, virtual or real objects. Dark color values in these displays become progressively more transparent such that a pure black is completely invisible. As a result, depth from shading cues are less reliable and the visual position of rendered objects in space becomes ambiguous. 


The inability to produce dark color values with fidelity in OST AR is especially problematic for rendering cast shadows. Manabe et al. \cite{Manabe:2018:CVS, Manabe:2019:SII} developed methods to produce more perceptually valid shadow rendering techniques for OST devices. Ikeda et al. \cite{Ikeda:2020:SIO} used simultaneous contrast to create the illusion of dark color values within the umbra of a shadow by applying brighter color values outside of the shadow's penumbra. In their evaluation, they found that their illusory shadow method caused the umbra of the shadow to be perceived as dark--despite the fact that nothing was rendered within the shadow's umbra itself. However, research on shadow generation algorithms gives rise to another question: how much fidelity is needed for accurate spatial perception in AR? Adams et al. \cite{Adams:2021:SLC} attempted to answer this question by evaluating surface contact judgements for both perceptually motivated shadow rendering methods, including a simple, dark gray shadow as well as a variant of the method developed by Manabe, Ikeda, and colleagues \cite{Manabe:2019:SII}, and a non-photorealistic rendering (NPR) method for shadows. Surprisingly, they found that more realistic rendering techniques for shadows did not inherently benefit surface contact judgements. Further, people's judgements for the NPR method, which produced a white shadow, were more accurate in both OST AR and VST AR head-mounted displays. 




Shadows may play an important role in alleviating inaccurate depth perception in augmented reality. 
As such, several evaluations on the influence shadow position and object shape on depth judgements have been conducted in both mobile AR \cite{ Berning:2014:SDP, Sugano:2003:ESR} and in OST AR \cite{Ping:2020:ESM, Gao:2019:IVO, Diaz:2017:DDP, Singh:2017:EFD, Hertel:2021:ARM}. 
The presence of cast shadows can dramatically improve the accuracy of people's judgements in perceptual matching tasks in OST AR displays \cite{Berning:2014:SDP, Diaz:2017:DDP, Gao:2019:IVO,Hertel:2021:ARM}. In perceptual matching paradigms, a person is asked to align a virtual object with a real world marker, or referent, that is positioned at a distance away from the viewer. As a result, perceptual matching may be considered a relative, or exocentric, measure of distance perception. 

In a recent study, Gao et al. \cite{Gao:2019:IVO} revealed that the shape of a shadow did not affect the accuracy of people's relative depth judgements in a perceptual matching task with floating, virtual targets. In the same study, Gao and colleagues also provided evidence that lighting misalignment can adversely affect people's accuracy when making relative depth judgements. However, prior research has found that ``drop shadows,'' in which shadows are rendered immediately below an object, regardless of the lighting in the environment around them, are an exception \cite{Berning:2014:SDP, Diaz:2017:DDP}. Drop shadows improve people's accuracy in perceptual matching experiments, despite their unrealistic positioning. 


Yet the results of these prior experiments, when taken in isolation, are somewhat limited 
since all of the aforementioned studies that evaluated shadows only rendered them with targets that floated above the ground \cite{Berning:2014:SDP, Diaz:2017:DDP,Ping:2020:ESM, Gao:2019:IVO, Hertel:2021:ARM}. This is important because anchoring shadows and detached shadows are perceived differently \cite{Casati:2004:MIS, Casati:2019:VWS}.
Anchoring shadows are formed when an occluding object and its shadow are connected (See Figure~\ref{fig:ball-in-box}~(top)). And detached shadows  are formed when an occluding object and its cast shadow do not touch (See Figure~\ref{fig:ball-in-box}~(bottom)). 
For example, the shape of a shadow when detached needs to only approximately conform to its object's shape \cite{Mamassian:1998:PCS, Casati:2019:VWS}. Accordingly, it makes sense that Gao et al.'s experiment found no differences between differently shape shadows, since they only evaluated hard, soft and 'round' (incorrect shape) shadow conditions when the test object was a rotating cube. 

In contrast, an anchoring shadow may be more sensitive to mismatches between the contours of a shadow and its associated object \cite{Mamassian:2004:ISS, Casati:2019:VWS}. However, as noted in Do et al. \cite{Do:2020:EOS}, AR depth perception experiments rarely incorporate multiple object shapes. Accordingly, in their mobile AR study they evaluated how color and luminance interacted to affect depth perception across four objects: a cube, a sphere, the Stanford bunny, and a low polygon version of the Stanford bunny. Although they predicted that simple shapes (i.e., the cube and sphere)  would be more affected by color and luminance depth cues than complex shapes, their predictions were only partially supported. The sphere was more affected by luminance than any other shape, but the cube was not susceptible to either color or luminance. Their results are reasonable considering that spheres have previously proven more challenging to perceive in depth than more distinctly faceted shapes, like cubes or icosahedrons, in VR \cite{Powell:2012:VPV}.

\begin{figure}[ht]
        \begin{minipage}[t]{.5\linewidth}
        \centering
        \includegraphics[width=0.6\linewidth,keepaspectratio]{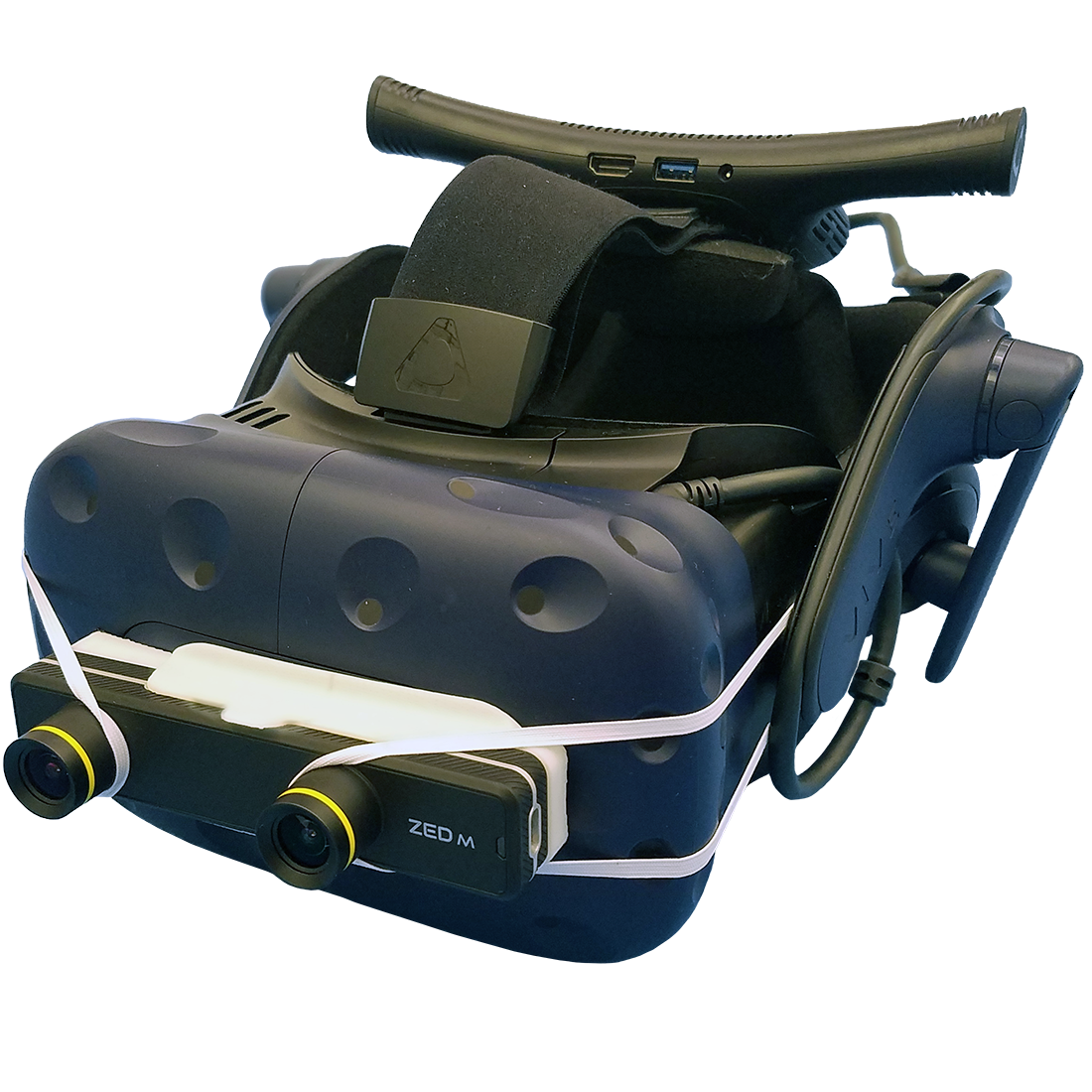}  
        \caption{A VR HMD with VST capabilities: the HTC Vive Pro with mounted Zed Mini camera}
        \label{fig:zed}
    \end{minipage} 
    \hfill
    \begin{minipage}[t]{.45\linewidth}
        \centering
        \includegraphics[width=0.6\linewidth, keepaspectratio]{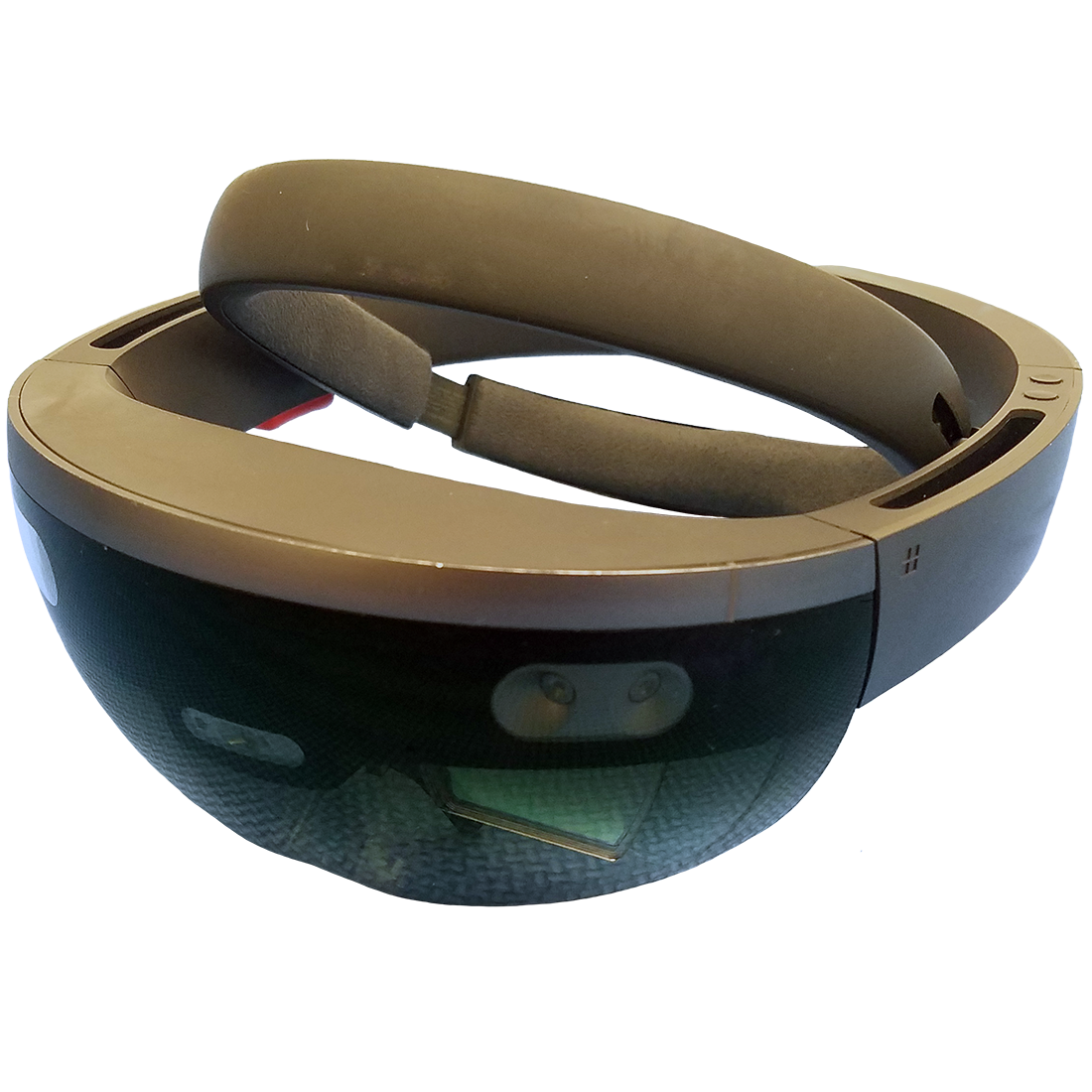}
        \caption{An OST HMD: the Microsoft HoloLens 1.}
        \label{fig:hololens}
    \end{minipage}
\end{figure}

\section{Overview} \label{overview}

We evaluated a viewer's sense of perceived surface contact over two studies, both of which were conducted on three immersive head-mounted display (HMD) types: a virtual reality (VR) display, a video see-through augmented reality display (VST AR), and an optical see-through augmented reality (OST AR) display. In both studies, shadows were manipulated to have either dark (photorealistic) or light (non-photorealistic) color values. In our first study, which is discussed in Section \ref{exp1}, we manipulated both shadow shading technique and object geometry to understand how the interplay between these two factors affected perceived surface contact in XR. In our second study (Section \ref{exp2}), we compared a simple geometric shape to a complex one, and we displayed them at different orientations. From the findings of our two studies, we were able to extrapolate XR developer and design guidelines, which are discussed in Section \ref{conclusions}. The following section (Section \ref{method}) discusses: our hardware setup (Subsection \ref{method_materials}), the software environment used for rendering to each display as well as the shaders used for rendering shadows (Subsection \ref{method_shadows}), and the equation used to displace test stimuli vertically based on visual angle (Subsection \ref{method_vertical}). 


\section{Methods} \label{method}

\subsection{Technical Setup} \label{method_materials}

The virtual reality environment was rendered using a wireless HTC Vive Pro (Figure \ref{fig:zed}), which has a maximum per eye resolution of $1440 \times 1600$ and an approximate field of view of  $110^\circ \times 113^\circ$. Position tracking for this condition was performed using the Vive's lighthouse tracking system.

The video see-through device also relied on the HTC Vive Pro for rendering. However, it created an augmented reality environment by combining virtual overlays with real time video feed, which was captured using the Zed Mini stereoscopic camera system. The Zed Mini was affixed to the front of the Vive Pro (Figure \ref{fig:zed}). The Zed Mini's camera feed restricted the HMD's resolution and field of view to $1280 \times 720$ and $90^\circ \times 60^\circ$, respectively. Position tracking was performed by the Zed Mini, which integrated its own native, inside-out tracking solution with the HTC Vive's tracking system.

The optical see-through environment was rendered using the Microsoft HoloLens 1 (Figure \ref{fig:hololens}). The OST display is unique since it relies on additive light to render images and is therefore unable to render dark color values with fidelity. Instead, the darker a color value is, the more transparent it becomes until it turns completely transparent at black. The HoloLens has an approximate resolution of $1268 \times 720$ and  field of view of $30^{\circ} \times 17^{\circ}$. Although the augmented field of view (FOV) of the HoloLens is narrow, outside of this viewing area users' vision is not occluded by the device.  For our experiment, position tracking was performed using the HoloLens' native inside-out tracking solution. 

 For the VST AR condition, virtual target objects were positioned within the real world environment using a Vive tracking puck before the start of each experiment. Similarly, for the OST AR system, this position calibration involved placing a HoloLens spatial anchor at the same predetermined position. This step was unnecessary for the VR condition since both the environment and the stimuli were virtual. A wireless computer mouse was employed for user inputs for all devices. Figure \ref{fig:participant} shows a participant wearing the wireless Vive Pro setup with the Zed Mini stereoscopic camera plugged in for the VST AR condition. The participant looks toward the spot on a table in front of them where the virtual target is rendered while holding a mouse. 

\subsection{Physical \& Virtual Environment}

In contrast to the two augmented reality display conditions, the virtual reality condition required a virtual reconstruction of the real world environment used for testing. Figures of both the physical and virtual testing environments can be seen in Figure \ref{fig:participant}. The virtual environment was created using photographed images of the real testing space along the walls and custom 3D models designed to match important furniture and foam floor tiles along the ground. In both the real and virtual environments, a table, on which experimental stimuli were displayed, was placed in the middle of the room. A disposable, azure blue cloth was draped along the top of the table with white tape wrapped around its edges. The tape was included to improve the table’s salience for the two AR systems, which relied on inside-out tracking.

\subsection{Software \& Shadows}\label{method_shadows}
Applications for each device were developed in Unity version 2017.4.4f1 with the C\# programming language. A virtual, directional light was positioned within each program so that a target object's shadow would lie behind and to the right of the object. To accomplish this, the orientation of the virtual light was set to $141^\circ$ along the x axis and $-141^\circ$ along the y axis. The color of the light had a slightly yellow tint with an RGB color value of (255, 244, 214). Shaders to render hard shadows were programmed using a variant of the HLSL language that is compatible with the Unity game engine. The shaders used to render target objects and their cast shadows used RGB color values to specify color output. All target objects were rendered with a middle gray RGB color value of 128. However, the cast shadow shader was developed to render shadows with custom color values. For the current study, we used grayscale RGB color values of 208 and 48 for the light and dark shadow shading methods, respectively. Visually, this created an off-white or lightly shaded appearance for the non-photorealistic (or light) shadow. And it rendered a dark gray colored shadow for the photorealistic (or dark) shadow. Because target objects were rendered with a middle gray color value of 128, there was a difference of 80 RGB color values for the dark (128-80=48) and the light (128+80=208) values used to inform the cast shadow shaders. Across all three devices, the same scripts were used for both object and cast shadow shading.  The equivalent values in CIE L*a*b* color space, which more accurately represents differences in lightness in perceptual space, are: $48=[19.9,0,0]$, $128=[53.6,0,0]$, and $208=[83.5,0,0]$.  


 The photorealistic, dark shadow condition was the most traditional method given it rendered shadows that adhered closely to the principle that shadows represent the absence of light. For most devices it created a perceptually valid impression of a shadow. In contrast, the non-photorealistic, light shadow condition added white light to generate a shadow instead of subtracting light. This created a both perceptually and photometrically incorrect cast shadow. Depictions of the shadow conditions can be seen in Figures \ref{fig:exp1_all}, \ref{fig:exp2_cubes}, and \ref{fig:exp2_dragons}.

\subsection{Vertical Displacement}\label{method_vertical}


Participants in our experiments judged whether targets were in contact with a surface; therefore, target stimuli had to be presented both on and above a surface for discrimination. We displaced objects vertically based on a viewer's visual angle to target. Given that participants conducted the study while seated, the eye height of the viewer was calculated by summing the average eye height of a person while seated and the height of the seat of the chairs used in our setup \cite{Harrison:2002:CSS}. Using the average eye height of a viewer, denoted as $h_e$, and the distance to a given target, $d_t$, we were able to solve for a series of three triangles from which we could extract the degree of vertical displacement, $d_v$, for target objects placed above surfaces. The trigonometric formulas used for this calculation are shown in the equations below:


\begin{equation} \label{eq:1}
\sigma =\tan^{-1}\left(\frac{d_t}{h_e}\right)  + \omega
\end{equation}

\begin{equation} \label{eq:2}
d_v = \left( \frac{ \tan(\sigma) \; {h_e} - d_{t}} {\tan(\sigma)} \right) 
\end{equation}

\begin{figure}[h]
    \centering
        \includegraphics[width=0.85\linewidth, keepaspectratio]{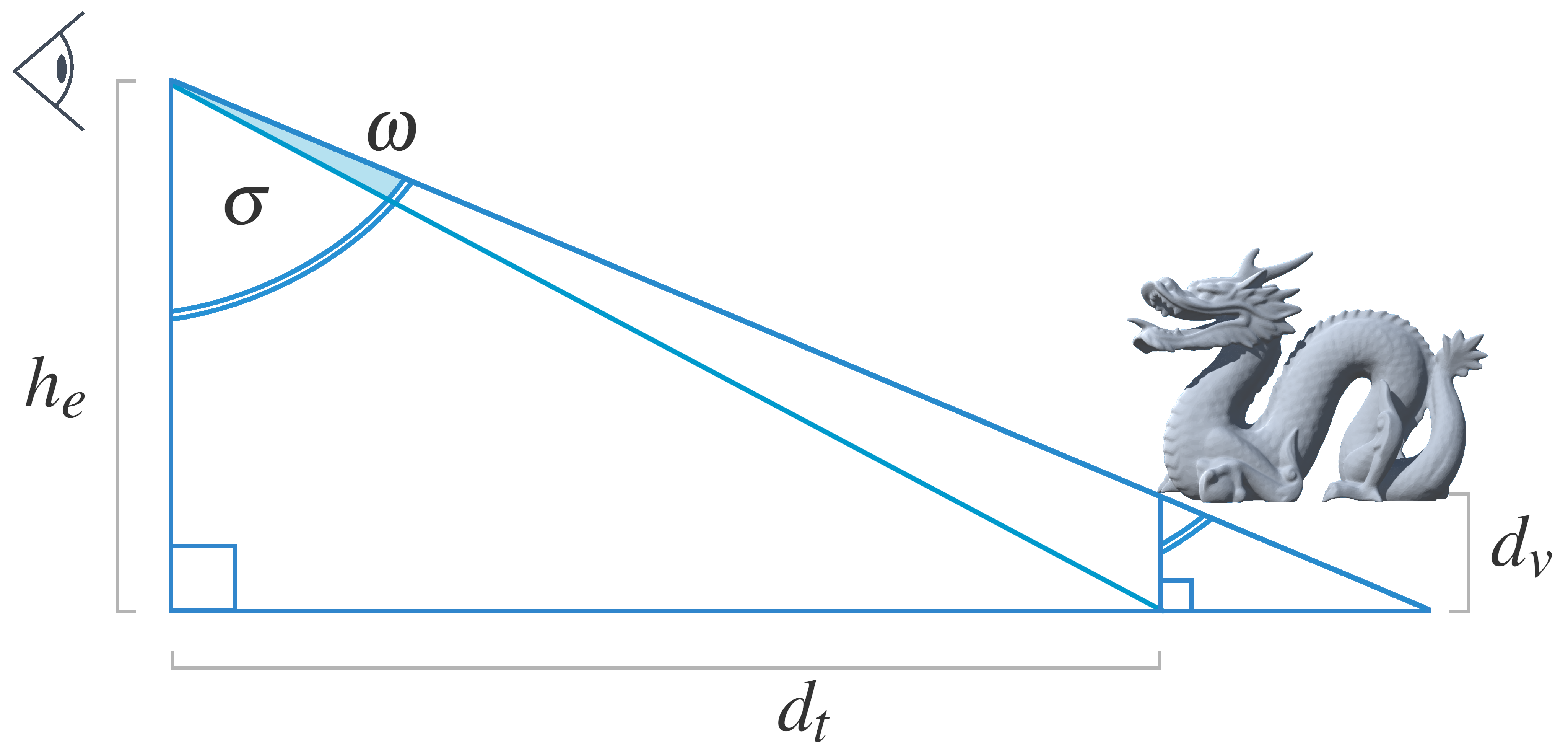}
        \caption{Visual depiction of trigonometric solution for vertical displacement.}
        \label{fig:angles}
\end{figure}

For equations \ref{eq:1} and \ref{eq:2}, $\omega$ represents the degree to which viewing angle was modified--in our case in increments of $0.6^{\circ}$--and $\sigma$ represents the updated viewing angle to the vertically displaced target object. The viewing angle of $0.6^{\circ}$ was selected during preliminary testing, where the value resulted in a sufficiently challenging degree of displacement for the experimental task.  Figure \ref{fig:angles} shows each variable in context for clarity. 

\section{Experiment 1: Shape and Shadow} \label{exp1}

In our first experiment, we evaluated how traditional (dark) and non-photorealistic (light) shading techniques for cast shadows affect one’s ability to perceive surface contact for a cube, an icosahedron, or a sphere (Figure \ref{fig:exp1_all}). These three objects were selected for their distinctive geometric properties.  In particular, the shape, shading, and shadows of the cube and sphere are recognizable and different. The icosahedron represents an in-between case, with its indistinct shape, but shading closer to that of the sphere and shadow closer to that of the cube. Our hypotheses for this experiment are as follows:

\textbf{H1}: We predicted that people’s ability to correctly perceive ground contact would be affected by an object's geometry, especially along the bottom edge that would be in contact with a surface. Specifically, we anticipated that people's likelihood of correct response would be lower for the sphere across all devices.  A perfect sphere has only a single point of contact with a ground surface and that singular point is occluded when viewed from above. In contrast, both the icosahedron and the cube were flat along the edge and benefited from many points of contact with the surface beneath them, although the surface area of the bottom of the icosahedron was smaller than that of the cube.  

\textbf{H2}: We anticipated that people would be more likely to correctly perceive surface contact when presented with the light shadow shading method, in comparison to the traditional dark shading method in both AR devices. Although, our prediction may seem counterintuitive, some prior research has shown that non-photorealistic shadows may function as well as photorealistic shadows as a depth cue for spatial perception \cite{Kersten:1997:MCS,Adams:2021:SLC}. 

\textbf{H3}: We also anticipated that there would be interactions between shape and shadow shading methods. However, \emph{a priori} we were uncertain how these interactions would be expressed. 


\subsection{Participants}

Six individuals in total (3M, 3F) aged 23--30 volunteered to participate in our first experiment, which used a well-founded psychophysical paradigm. Psychophysics is a class of psychological methods that quantitatively measures perceptual responses to changes in physical stimuli \cite{Farell:1999:VRP}. These methods can use a small number of participants to make a large number of simple, behavioral responses that reveal underlying perceptual processes, gaining experimental power from a large number of observations. Psychophysical paradigms have proven highly replicable and robust since they employ judgments or adjustments with low individual variance \cite{Epstein:1980:SOB,Smith:2018:SIB}.   Psychophysics methods have previously been employed successfully by other XR research groups \cite{Bulthoff:2001:VAV,Schuetz:2019:EFL,Frankenstein:2020:APA, Madison:2001:UIS}. 

All participants had normal or corrected-to-normal vision. Our experimental methods were approved by the local institutional review board, written consent was obtained from all subjects prior to participation, and each participant was paid 20 USD for 2-3 hours of their time.

\begin{figure}[b]
    \centering
        \includegraphics[width=0.85\linewidth, keepaspectratio]{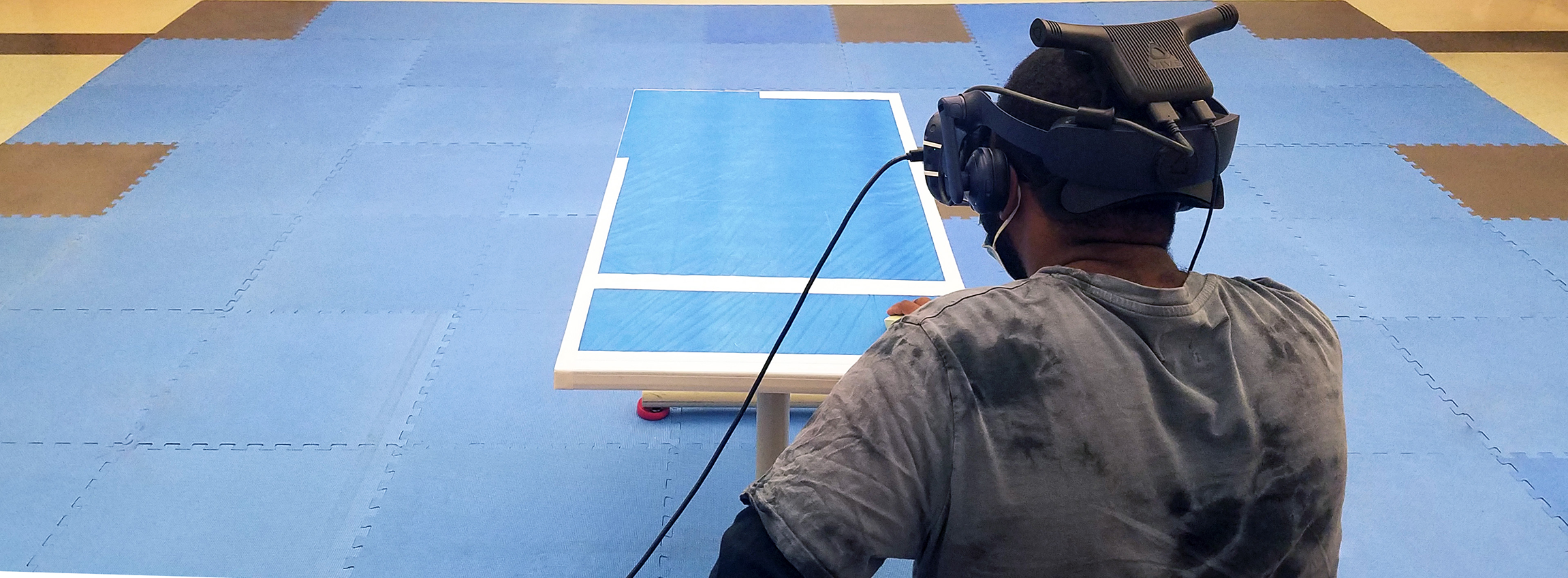} 
        \includegraphics[width=0.85\linewidth, keepaspectratio]{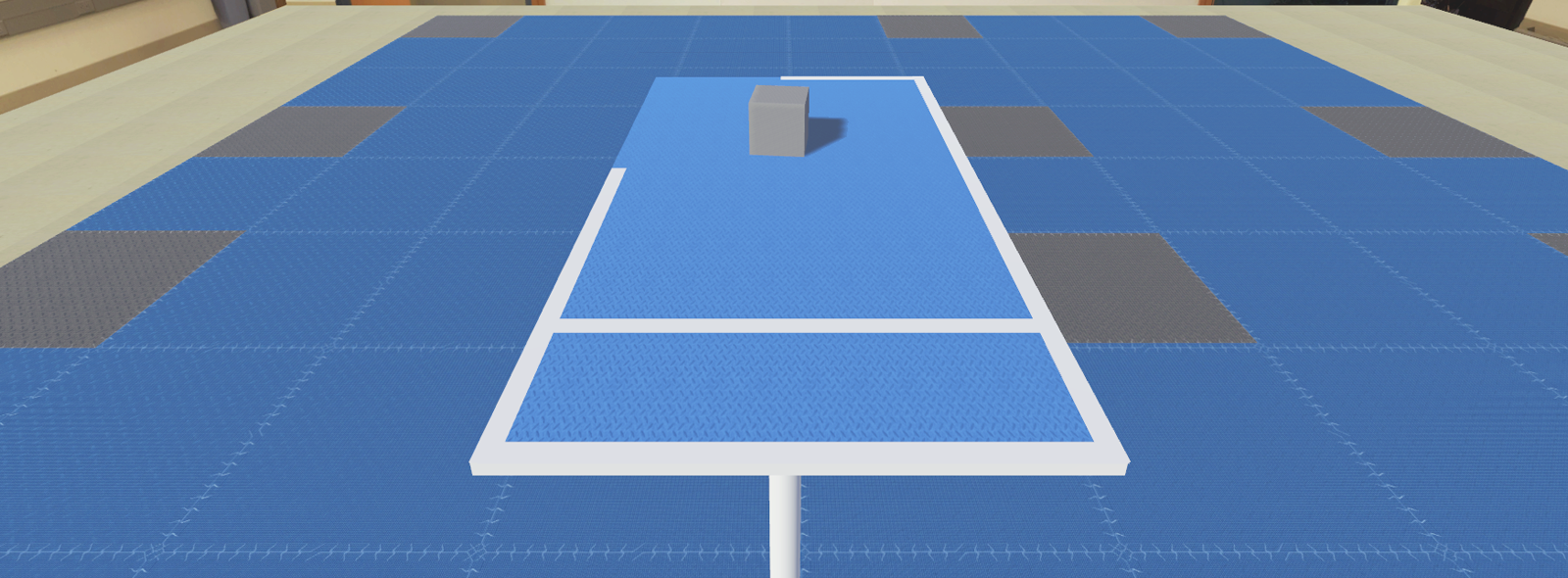}
        \caption{A participant views the experiment in the VST AR condition (top). An image of the virtual environment used for the VR condition (bottom). }
        \label{fig:participant}
\end{figure} 

\subsection{Design}

To address our hypotheses, we utilized a 3 (XR display) $\times$ 3 (target shape) $\times$ 2 (shadow shading)  $\times$ 6 (target height) within-subjects design to evaluate the effects of object shape and shadow shading on surface contact judgments for objects at different vertical displacements. Although we conducted our experiment with three different XR displays, we did not make any direct statistical comparisons across displays due to the high variance in optical and graphical properties across devices. However, we evaluated the same experimental conditions for each XR display. As such, this section describes the experimental design that was used for each XR device.


Each participant evaluated 3 object shapes with 2 different shadow shading conditions, which resulted in 6 unique combinations of experimental stimuli. The target shapes evaluated were a cube, an icosahedron, and a sphere. The shadows were shaded with either dark or light color values. Figure \ref{fig:exp1_all} displays all of the target shape and shadow shading conditions. All targets were rendered on a table 1 meter in front of the viewer (Figure \ref{fig:participant}). The size of the targets was adjusted to be 11.2 cm wide. Accordingly, the cube had a length, width, and height of 11.2 cm, the sphere's diameter was 11.2 cm, and the icosahedron was scaled so that its width was approximately 11.2 cm. 


We employed a temporal two-alternative forced choice paradigm with method of constant stimuli---a classic psychophysical method \cite{Farell:1999:VRP,Bogacz:2006:POD}---to evaluate people's perception of surface contact. Two-alternative forced choice, or 2AFC, is a method for assessing someone's sensitivity to a change in stimuli. It involves a \emph{forced choice}, because participants must chose between one of two options.  A temporal 2AFC paradigm may also be referred to as a two-interval forced choice (2IFC) paradigm. After viewing both stimuli in a sequence, participants in our experiment are then asked "Which object is closer to the ground?". Compared targets were always of the same shape and shadow condition for each trial. 


How the six levels of vertical displacement were presented was dictated by \emph{method of constant stimuli}. As such, at least one of the target stimuli was always placed in contact with the surface beneath it. The other object was then displaced vertically by one of six heights between 0 and 6mm at regular intervals of $.06^{\circ}$ changes in viewing angle. When displacement was 0, both the first and second targets were positioned in contact with the surface. 

Changes in vertical displacement were subtle. All six height displacements are displayed in Figure \ref{fig:exp1_all}, and the calculations for vertical displacement are described in detail in Section \ref{method_vertical}. Each height comparison was presented 20 times each, except for when both the first and second object heights were equal (at 0). In this case, the height conditions were only presented 10 times each. Stimulus pairs were presented pseudo-randomly such that there were no repetitions of shape $\times$ shadow $\times$ height combinations before all combinations were presented once. In addition, the order of presented stimuli was balanced so that the number of trials in which the target was presented in contact with the ground first and the number of times it was presented second were equivalent for each condition.

Trials for the task were blocked by device and order of device was counterbalanced across subjects. For each XR device, a participant completed 660 2AFC trials. Each participant thus completed 1,980 trials total, and 11,880 data points were collected across all subjects. We collected data over a large number of trials, which is common practice in psychophysical paradigms, to increase accuracy and ensure low variance in our study \cite{Prins:2016:PPI}. By evaluating a large number of simple behavioral responses on a small number of participants, this family of paradigms is better able to evaluate perceptual behaviors that have little variance between individuals \cite{Epstein:1980:SOB,Smith:2018:SIB}. We further confirmed that variance in our collected data was not due to between-participant variation by calculating the intra-class correlation coefficient (ICC), the calculation of which is reported in Section \ref{exp1_results} (Results).


\subsection{Procedure}

Before the experiment began, each individual was informed about the study and filled out a consent form. They were told that they may stop the experiment at any time and that they were allowed to take breaks during the experiment if needed. Both the experimenter and the volunteer wore face masks while maintaining 2 m of separation between each other during the study. 

The experiment was blocked by device.
As such, the participant picked up one of the three XR HMDs, which was predesignated based on a device ordering that was counterbalanced across participants, to begin the experiment.  They were then verbally instructed on how to interact with the system and they performed 10-20 practice trials that were randomly selected trials from the experiment. After the volunteer expressed that they were comfortable with the system and experimental paradigm, the experiment began. 

For the temporal 2AFC paradigm with method of constant stimuli, each  stimulus was presented for 600 ms. In between each stimulus pair, there was an interval of 800 ms in which a random pattern was presented at the position of the object to avoid visual aftereffects. After the presentation of each stimulus pair, the participant was asked ``Which stimulus is closer to the ground?'' They were told that at least one of the objects was positioned on the ground. They then responded using the left mouse button to indicate the first stimulus and the right button for the second stimulus.  The experiment was self-paced, and both the user’s response as well as their response time were recorded for each trial. The next trial began 1000 ms after the participant responded to the previous trial---unless the experiment was paused. After every 66 trials, participants were presented with a visual prompt that asked if they needed to take a break. 

Between devices, participants were also required to take a break to prevent fatigue. During the break, they filled out a brief survey that asked them to describe any strategies that they used to determine ground contact. After finishing the experiment in its entirety, they were asked to fill out a final survey. Although we did not directly compare peoples' performance across XR devices in this study, due to the significant differences in display properties, in the final survey we asked participants about their experiences in each device. We believed this information would be informative for interpreting our results. Therefore, in the post-experiment survey, they were asked to rank the difficulty of the experimental task for each device and to rank the displays' quality of graphics. They were also asked if their strategy for determining ground contact differed across devices.


\begin{figure*}[hb!]
    \centering
        \includegraphics[width=0.325\linewidth, keepaspectratio]{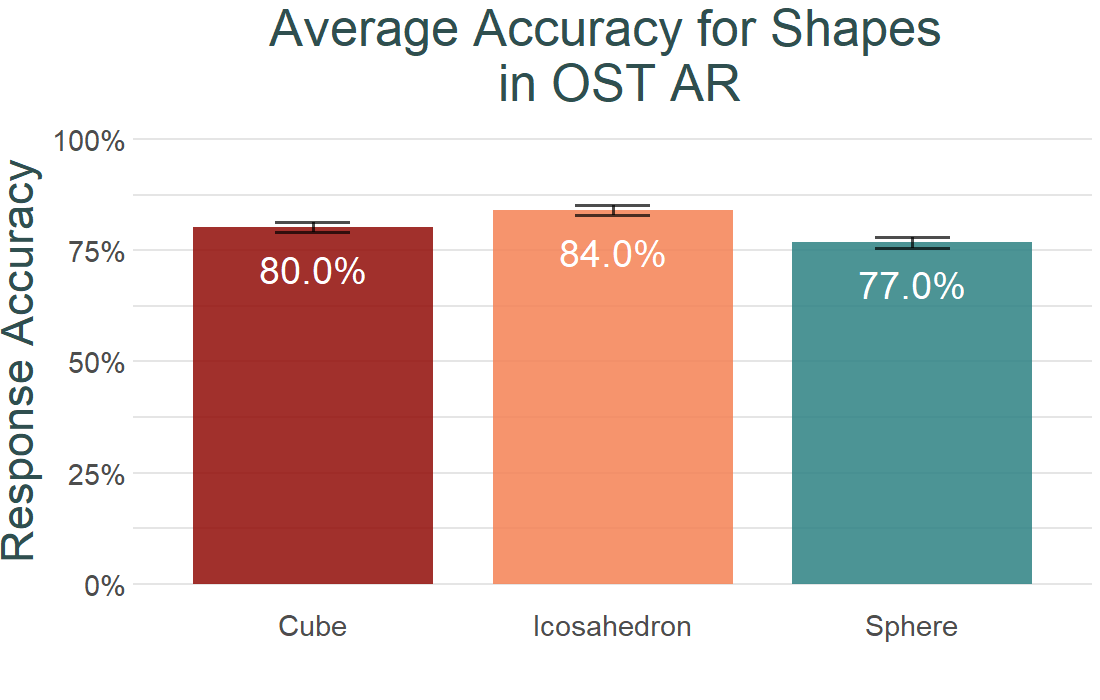}
        \includegraphics[width=0.325\linewidth, keepaspectratio]{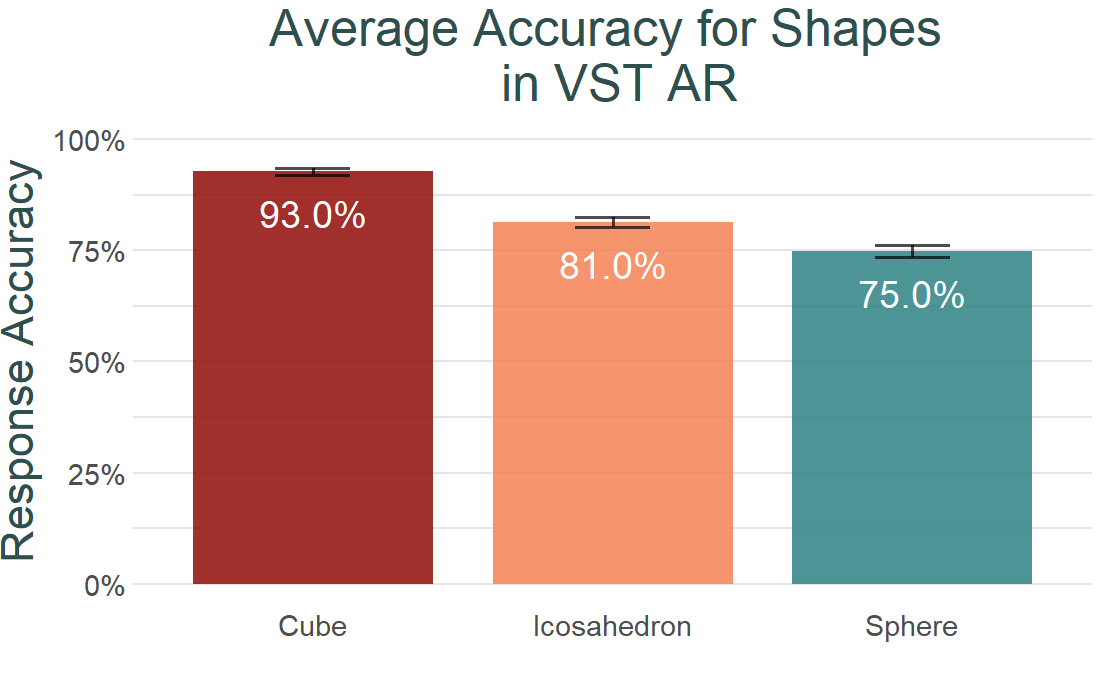}
        \includegraphics[width=0.325\linewidth, keepaspectratio]{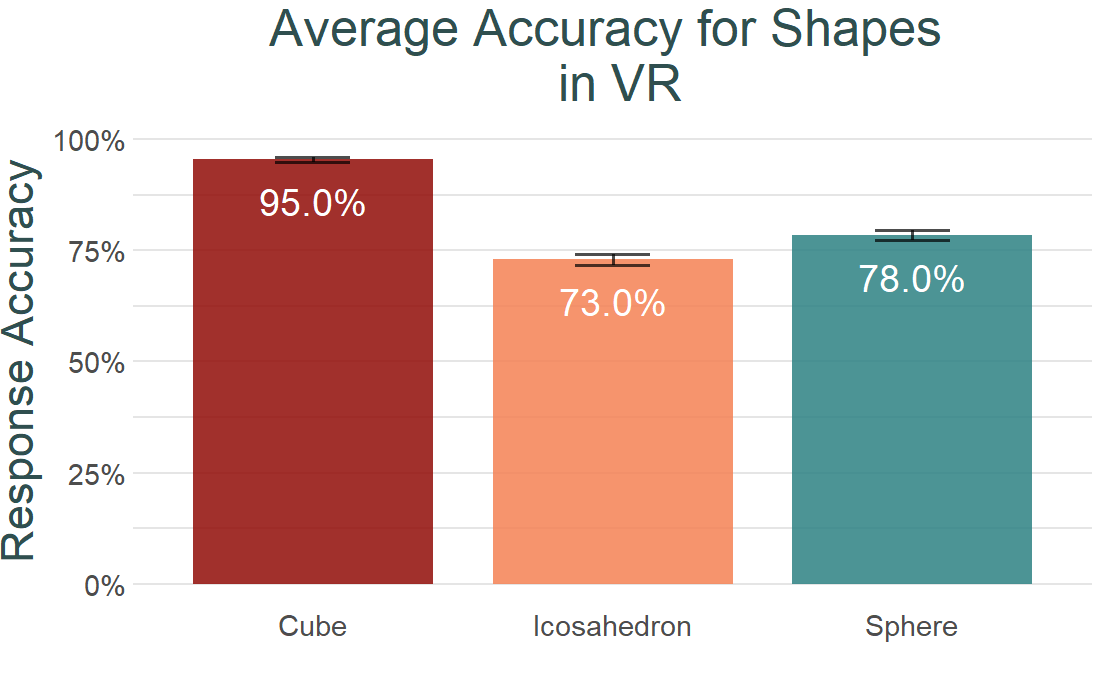}
        \caption{Experiment 1 --- Average percentage of correct responses for each target object shape by display condition: optical see-through AR (left), video see-through AR (center), and virtual reality (right). The effects of shape on surface contact judgement were complex, with significant but different effects of target shape for each display. However, people's judgements with the cube were significantly better than judgements with the sphere in all three devices. }
        \label{fig:exp1_shape_results}
\end{figure*}

\subsection{Results} \label{exp1_results}

Our experiment used a small-N design. Therefore, before performing additional statistical analyses we confirmed that variance in our collected data was not due to between-participant variation by calculating the intra-class correlation coefficient (ICC). The ICC measures the amount of variance accounted for by a grouping variable. For our analysis, individual participant was selected as the grouping variable, where an ICC value of 1 indicated  that  any  variance  in  the  data  is  between-participants  while  a  value  of  0  indicated  that  no  variance  in  the data is due to between-participant factors. We found that $\tau_{ost}$ = .009 for data collected with the OST AR display, which indicated negligible variance was caused by between-participants factors. Similarly low ICC values were found for the VST AR  and VR display data with ICCs of $\tau_{vst}$ = .012 and $\tau_{vr}$ = .042, respectively.

\paragraph{Statistical Analyses} 
Participants were asked to make binary decisions about which object was positioned closer to the ground (first or second object) in a 2AFC task. Therefore, for our statistical analyses we used binary logistic regression models, which are appropriate for dichotomous outcome variables, to analyze participants' judgments. We used the glm function from the stats package in R \cite{Ihaka:1996:R} to conduct logistic regressions by specifying binomial errors and a logit link function. Because we wanted to analyze people's perception in each device, we ran separate models for each of the XR displays. For each display, we modeled binary outcomes (accuracy: correct (1) or incorrect (0)) for our predictors: object shape, shadow shading, and height. Height was recorded in millimeters then centered at zero and treated as continuous. Object shape (3 levels: cube, icosahedron, sphere) and shadow shading (2 levels: dark, light) were treated as categorical factors. Interactions between shape and shadow were included in the models.

We used these models to test three planned comparisons for each device to understand how object shape and shadow influenced people's surface contact judgments. These comparisons were: 1) whether surface contact judgments differed for different object geometries (i.e., \textbf{H1}~the main effect of object shape); 2) the difference in surface contact judgments between dark and light shadow shading across all other conditions (i.e., \textbf{H2} the main effect of shadow shading); and 3) interactions between object shape and shading (i.e., \textbf{H3}).


In order to examine whether there were main effects, we coded shadow and shape factors using deviation coding (also known as effect or sum coding). For the shape factor, the sphere was set as the reference group (i.e., coded as -1). For the shadow factor, the light shadow condition was coded as .5 and dark shadow was coded as -.5. Using this deviation coding also allowed us to observe whether there was a main effect of height. The general logistic regression equation is depicted in Equation \ref{eq:glm_model} below.

\begin{equation}
   \begin{split}
    \log\left(\frac{p}{1-p}\right) = B_0 & + B_1(\mbox{shadow}) + B_2(\mbox{cube}) \\ & + B_3(\mbox{icosahedron}) + B_4(\mbox{height}) \\
    & + B_5(\mbox{shadow} \times \mbox{cube}) \\
     & + B_6(\mbox{shadow} \times \mbox{icosahedron})
   \end{split} \label{eq:glm_model}
\end{equation}

For the shape variable (\textbf{H1}), we were interested in whether the sphere lead to lower accuracy than the other two shapes, collapsed across shadow and height. In order to answer this question, we conducted planned contrasts on the aforementioned model comparing each shape to one another, using the Bonferroni correction to account for multiple comparisons. 

In order to examine whether there was a main effect of shadow shading (\textbf{H2}), the shadow regression coefficient represents the difference between dark and light shadows averaging over shape and height. We were also interested in a potential interaction between shape and shadow (\textbf{H3}). Specifically, was each shape affected by the shadow shading? In order to answer this question, we calculated the shadow simple slopes for each shape. 

For the sake of simplicity, we discuss the implications of our findings as well as which factors are significant in text, and we report the full details of the effects of factors in our analyses in Table \ref{tab:exp1_results}.

\begin{figure*}[ht]
    \centering
        \includegraphics[width=0.325\linewidth, keepaspectratio]{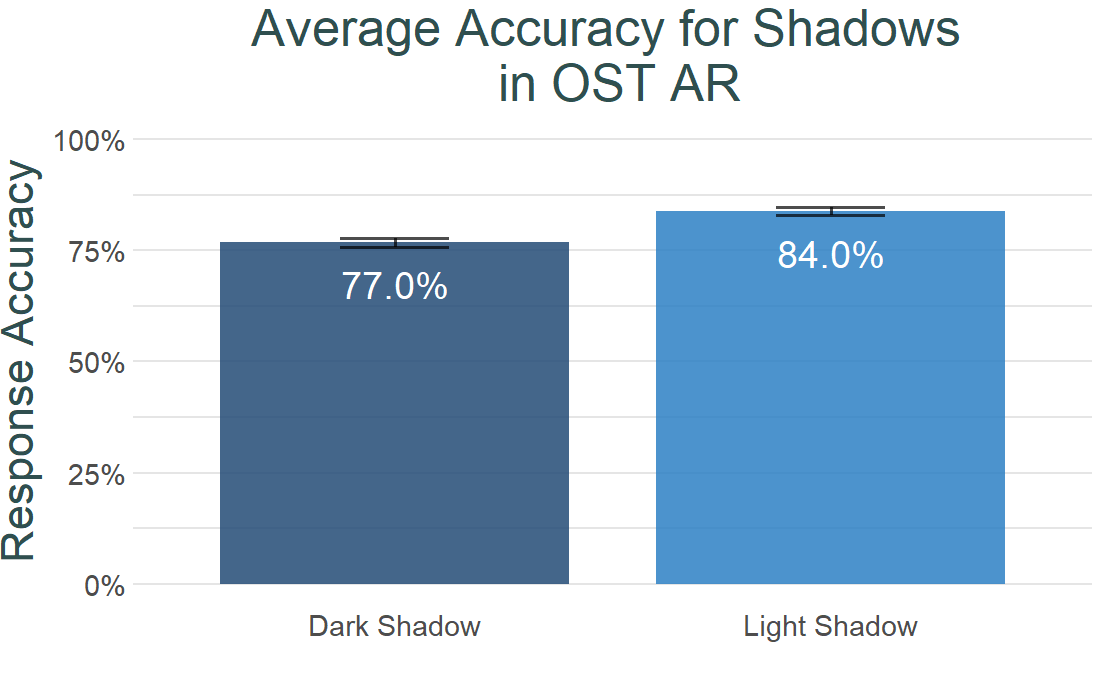}
        \includegraphics[width=0.325\linewidth, keepaspectratio]{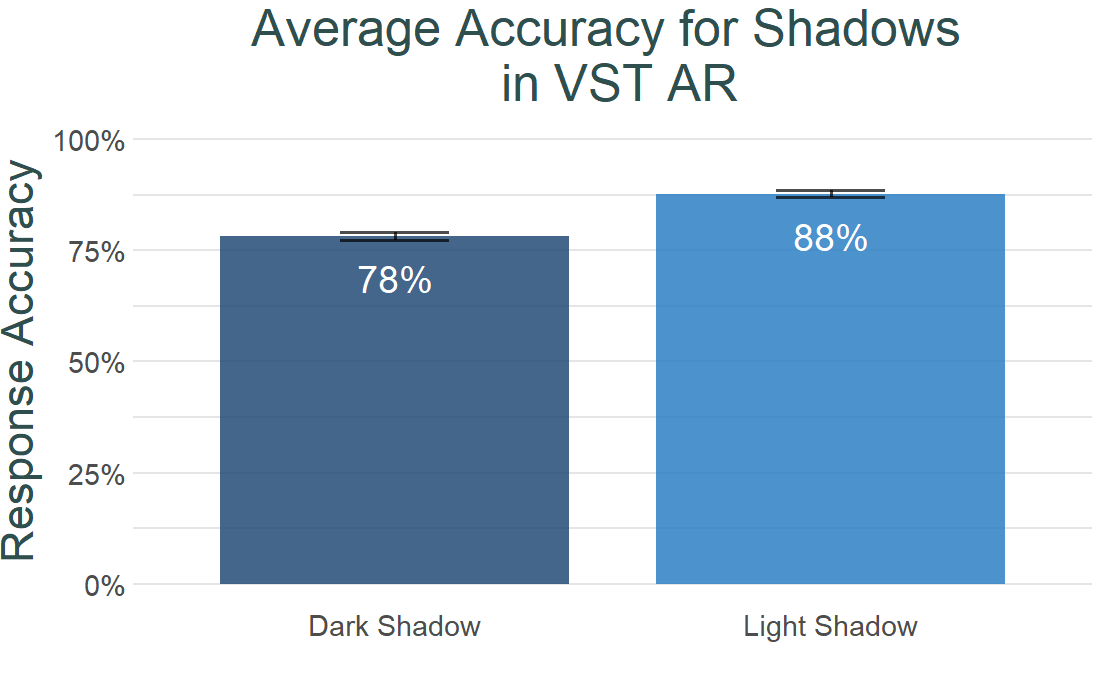}
        \includegraphics[width=0.325\linewidth, keepaspectratio]{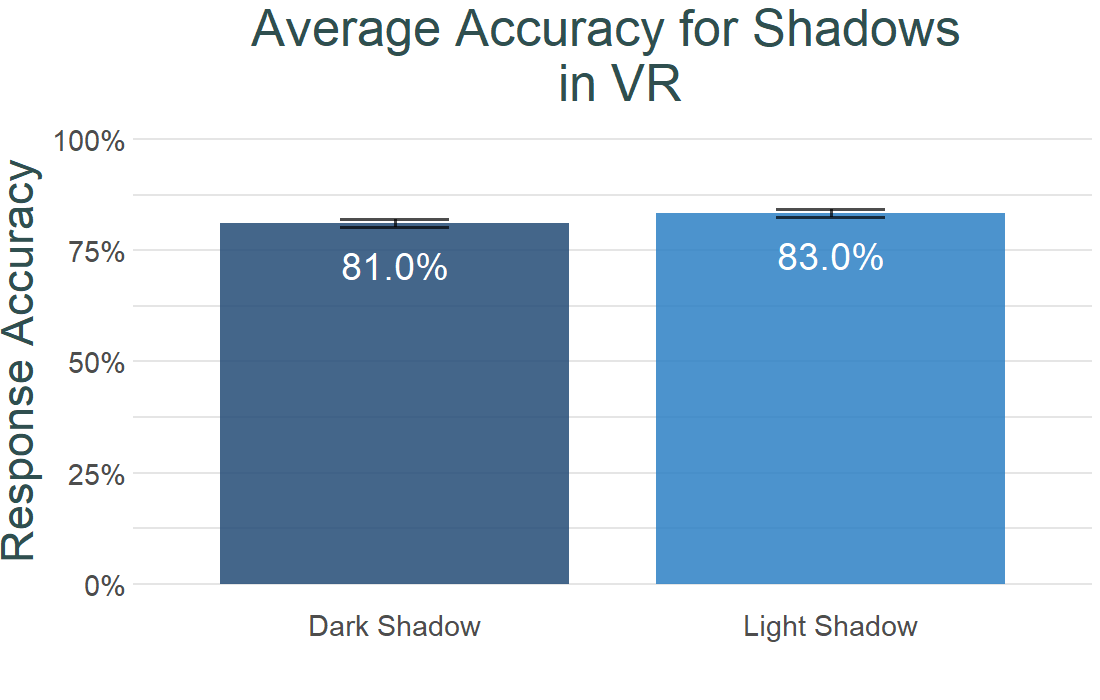}
        \caption{Experiment 1 --- Average percentage of correct responses for each target object shadow by display condition: optical see-through AR (left), video see-through AR (center), and virtual reality (right). People were significantly more accurate when shadow shading was light, rather than dark in both AR device conditions. }
        \label{fig:exp1_shadows_results}
\end{figure*}

\subsubsection{Optical See-through AR}


We show the average percent of correct response for each of the evaluated main effects in Table \ref{tab:exp1_acc}.
The results of our statistical analyses are reported in Table \ref{tab:exp1_results}.  We expected participants to be more likely to correctly indicate which object was on the ground as the height of target objects increased. The improvement of participants' performance as height increased is demonstrated statistically by a main effect of height in our logistic regression model ($OR = 1.09$, $p < .001$). Collapsed across all shape and shadow conditions, the odds ratio ($OR$) of 1.09 indicates that for every 1mm increase in height, the odds of correctly stating which object was closer to the ground increased by a factor of 1.09. 


\paragraph{H1: Does object shape matter?}

People were least accurate when assessing surface contact given the sphere, with 77\% accuracy on average. In comparison, people were 80\% accurate when presented with the cube and 84\% accurate when presented with the icosahedron (See Figure~\ref{fig:exp1_shape_results}).
In our paired comparisons of people's accuracy for each target shape, when collapsed across shadow and height, we found that people's accuracy when they were shown the sphere was significantly worse than their accuracy when shown the cube~($OR = 1.42$, $p < .01$) or the icosahedron~($OR = 1.67$, $p < .001$).  
In other words, participants were 1.42 times or 1.67 times more likely to choose the correct object with the cube or icosahedron, respectively, than with the sphere. 
There was not a significant difference in response accuracy between the cube and icosahedron~($OR = .85$, $p = .50$). 

\paragraph{H2: Does shadow shading method matter?} 
People were 77\% accurate when assessing surface contact when presented with a dark shadow and 84\% accurate when presented with a light shadow (See Figure~\ref{fig:exp1_shadows_results}). 

The main effect for shadow ($OR = 1.71$, $p < .001$) indicates that a correct response was 1.71 times more likely when the object was presented with a light shadow compared to a dark shadow.

\paragraph{H3: Is there an interaction between shape and shadow?} 
Analysis of the shadow by shape simple slopes indicated that the effect of shadow was only significant for the cube ($OR = 4.21$, $p < .001$), a finding that indicates that people were more 4.21 times more likely to make a correct response when the cube was rendered with a light shadow. The predicted probabilities of correct response for each shape and shadow are displayed in Figure \ref{fig:exp1_slopes} (left).
On average,  when presented with the cube, people were 70\%~($SE=1.9\%$) accurate with the dark shadow was presented and they were 90\%~($SE=1.2\%$) accurate with the light shadow.

\begin{table}[t]
\caption{
Experiment 1 --- The accuracy for each shape and shadow condition tested reported with the standard errors. } 
\label{tab:exp1_acc}
\centering
\resizebox{0.96\columnwidth}{!}{%
\begin{tabular}{lllcc}
\multicolumn{5}{c}{\textbf{Accuracy for Experiment 1}} \\ 
\toprule
           Condition &           &\multicolumn{1}{c}{OST AR}   &\multicolumn{1}{c}{VST AR}   &\multicolumn{1}{c}{VR}  \\
           \\ 
     Cube &           &80\% $\ [1.2\%]$  &93\% $\ [\ \ .7\%]$  &95\% $\ [\ \ .6\%]$  \\
     Icosahedron &    &84\% $\ [1.1\%]$  &81\% $\ [1.1\%]$  &73\% $\ [1.3\%]$  \\
     Sphere &         &77\% $\ [1.2\%]$  &75\% $\ [1.3\%]$  &78\% $\ [1.0\%]$  \\
\midrule \\
     Dark &    &77\% $\ [\ \ .1\%]$  &78\% $\ [1.0\%]$  &81\% $\ [\ \ .9\%]$  \\ 
     Light &   &84\% $\ [\ \ .1\%]$  &88\% $\ [\ \ .8\%]$  &83\% $\ [\ \ .9\%]$  \\
\bottomrule
\end{tabular}}%
\end{table}

\subsubsection{Video See-through AR}

The average probabilities of correct response for each main condition are shown in Table \ref{tab:exp1_acc}, and the results of our statistical analyses are reported in Table~\ref{tab:exp1_results}.  For VST~AR we again found a main effect of height ($OR = 1.07$, $p < .001$). 
Accordingly, as the height of the vertically displaced object increased, participants were more likely to correctly indicate which object was on the ground. 

\paragraph{H1: Does object shape matter?}
People were most likely to produce a correct response when presented with the cube with 93\% accuracy, on average, and they were least likely to make a correct response when presented with a sphere with 75\% accuracy. People's accuracy for the icosahedron was 81\% on average (Figure~\ref{fig:exp1_shape_results}). 
Planned contrasts for the shape variable indicated that the odds of providing a correct response significantly differed between all shape comparisons. The cube was more likely to yield a correct response than either the sphere ($OR = 4.90$, $p < .001$) or the icosahedron  ($OR = 2.93$, $p < .001$). And the odds of providing a correct response for the icosahedron was higher than the odds for the sphere ($OR = 1.67$, $p < .001$).

\paragraph{H2: Does shadow shading method matter?} 
The main effect for shadow ($OR = 2.28$, $p < .001$) indicates that a correct response was 2.28 times more likely when the object was presented with a light shadow compared to a dark shadow. Overall, people were 88\% accurate on average when presented with the light shadow and 78\% accurate when presented with the dark shadow (See Figure~\ref{fig:exp1_shadows_results}). 

\begin{figure*}[t]
    \centering
    \includegraphics[width=\linewidth, keepaspectratio]{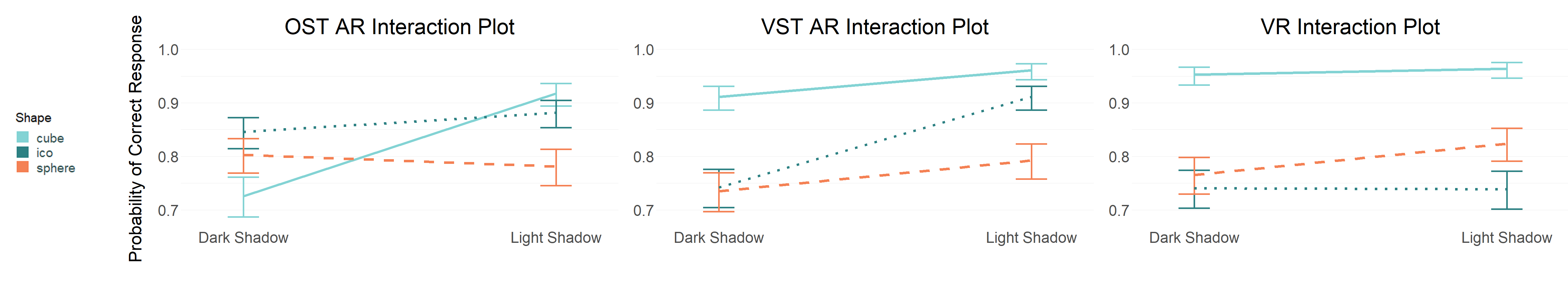}
        \caption{
        Experiment 1 --- Predicted probability of correct response for shape (cube, icosahedron, sphere) and shadow (dark, light) interactions  with 95\% confidence. In VST AR, all shape x shadow interactions were significant. In contrast, for OST AR the effect of shadow was only significant for the cube and for VR it was only significant for the sphere.   }
       \label{fig:exp1_slopes}
\end{figure*}

\paragraph{H3: Is there an interaction between shape and shadow?} 
Analysis of the shadow by shape simple slopes indicated that the effect of shadow was significant for all three shapes. That is, the probability of providing a correct response was higher when a light shadow was presented regardless of whether the object presented was a sphere ($OR = 1.38$, $p < .05$), a cube ($OR = 2.40$, $p < .001$), or an icosahedron ($OR = 3.58$, $p < .001$). The predicted probabilities of correct response for each shape and shadow are displayed in Figure \ref{fig:exp1_slopes} (center).
The average accuracy for people's judgments with the cube was 90\% $(SE=1.2)$ on average for the dark shadow condition and 95\% $(SE=0.8)$ for the light shadow condition. For the icosahedron, people were 73\% $(SE=1.8)$ accurate with the dark shadow and 90\% $(SE=1.2)$ accurate for the light shadow. For the sphere people were 72\% $(SE=1.8)$ accurate with the dark shadow and 78\% $(SE=1.8)$ accurate for the light shadow.



\newcolumntype{C}{@{\extracolsep{3cm}}c@{\extracolsep{0pt}}}%
\sisetup{add-integer-zero=false}

\begin{table*}[t]
\caption{Experiment 1 --- Results of planned comparisons using binary logistic regression models for each display condition are displayed. \emph{B} is the regression coefficient, $SE_{B}$ is the standard error of the regression coefficient, $OR$ is the odds ratio, and $CI_{OR}$ is the confidence interval associated with the odds ratio. Negative values for \emph{B} indicate that the first factor in the comparison was more accurate, whereas positive values indicate that the second factor was more accurate. } 
\label{tab:exp1_results}
\resizebox{\linewidth}{!}{%
\begin{tabular}{ll SlSl SlSl SlSl}
\\
\multicolumn{14}{c}{\textbf{Results for Experiment 1}} \\
\toprule  
\\
&& \multicolumn{4}{c}{\hspace{22pt}{OST AR}}  &\multicolumn{4}{c}{\hspace{22pt}{VST AR}}  &\multicolumn{4}{c}{\hspace{22pt}{VR}} 
\\ \\
    Predictor  &
    &\emph{B} &$SE_{B}$  &$OR$ &\multicolumn{1}{c}{$[CI]_{OR}$}
    &\emph{B} &$SE_{B}$  &$OR$ &\multicolumn{1}{c}{$[CI]_{OR}$} 
    &\emph{B} &$SE_{B}$  &$OR$ &\multicolumn{1}{c}{$[CI]_{OR}$} \\
\cmidrule(lr{12pt}){1-2}  \cmidrule(l{22pt}r){3-6} \cmidrule(l{22pt}r){7-10} \cmidrule(l{22pt}r){11-14} \\ 
    Shape 
    &(cube vs ico)                     
        &.16    &.12   & .85   & $[\ \ .64,\ 1.13]$   
        &-1.08***   &.15   & 2.93   & $[2.06,\ 4.17]$  
        &-2.10***   &.16   & 8.20   & $[5.66,\ 11.9]$   \\ 
    &(cube vs sphere)                  
        &-.35$\ $**     &.11   & 1.42   & $[1.09,\ 1.85]$  
        &-1.59***   &.14   & 4.90   & $[3.51,\ 6.84]$  
        &-1.78***   &.16   & 8.20   & $[4.08,\ 8.67]$  \\
    &(ico $\ \ \ $vs sphere)                   
        &-.52$\ $***     &.11   & 1.67   & $[1.29,\ 2.17]$  
        &-.51$\ $***     &.11   & 1.67   & $[1.29,\ 2.17]$  
        & .32$\ $**  &.20   & 1.06   & $[\ \ .57,\ \ .92]$ 
    \\ \\
    Shadow 
    &(dark $\ $vs light)    
        & .54$\ $***   &.09    & 1.71  & $[1.43,\ 2.05]$  
        & .82$\ $***   &.11    & 2.28   & $[1.85,\ 2.83]$
        & .21          &.11    & 1.23   & $[\ \ .99,\ 1.54]$ 
    \\ \\      
    Shape$\times$Shadow
    &(cube:\quad dark vs light)    
        & 1.44$\ $***  &.17    & 4.21   & $[3.02,\ 5.85]$  
        & .88$\ $***  &.24    & 2.40   & $[1.49,\ 3.85]$
        &.27      &.28    &1.31   &$[\ \ .76,\ 2.27]$  \\
    &(ico: \quad $\ $ dark vs light)     
        &.31    &.16   &1.36    & $[\ \ .99,\ 1.88]$
        & 1.28$\ $***    &.17    & 3.58   & $[2.58,\ 4.98]$
        &-.01   &.13   & .99  & $[\ \ .76,\ 1.29]$    \\
    &(sphere: dark vs light)  
        &-.13   & .14     &.88     & $[\ \ .66,\ 1.16]$
        & .32$\ $*  &.14   & 1.38   & $[1.05,\ 1.81]$
        & .36$\ $*  &.14   & 1.43   & $[1.08,\ 1.90]$    \\ \\ 
    Height&                                  
        &.09$\ $*** &.01  &1.09   &$[1.08,\ 1.10]$
        &.07$\ $*** &.01  &1.07   &$[1.06,\ 1.08]$
        &.06$\ $*** &.01  &1.06   &$[1.05,\ 1.07]$ \\
    Intercept&                               
        &1.63$\ $*** &.05 &5.10  &$[4.62,\ 5.63]$ 
        &1.88$\ $*** &.06 &6.55  &$[5.87,\ 7.35]$ 
        &1.85$\ $*** &.06 & 6.36 &$[5.69,\ 7.16]$ \\ 
\bottomrule 
    &\multicolumn{9}{c}{ }   &\multicolumn{4}{c}{$* p < .5$ \qquad $** p < .01$ \qquad $*** p < .001$} \\
\end{tabular}}%
\end{table*}

\subsubsection{Virtual Reality}

The average rate of correct response for each main condition is reported in Table \ref{tab:exp1_acc}, and the results of our statistical analyses are reported in Table~\ref{tab:exp1_results}. For VR, we found a main effect of height ($OR = 1.06$, $p < .001$) such that for every 1mm increase in height, the odds of correctly stating which object was closer to the ground increased by a factor of 1.06. 

\paragraph{H1: Does object shape matter?} 
People were most likely to make a correct response when presented with the cube (95\% accuracy on average), and they were least likely to make a correct response when presented with the icosahedron (73\%). People's accuracy for the sphere was 78\% (Figure~\ref{fig:exp1_shape_results}). 
The probability of providing a correct response significantly differed for all shape comparisons. Specifically, planned contrasts, when collapsed across shadow and height, showed that participants were significantly more likely to make a correct response when presented with the cube than the icosahedron ($OR = 8.20$, $p < .001$) or sphere ($OR = 5.95$, $p < .001$). They were then more likely to make a correct response given the sphere over the icosahedron ($OR = .73$, $p < .01$). 

\paragraph{H2: Does shadow shading method matter?} 
The average accuracy for surface contact judgments was 81\% for dark shadows and 83\% for light shadows (See Figure~\ref{fig:exp1_shadows_results}). 
The main effect for shadow was not significant ($OR = 1.23$, $p = .07$), suggesting that, averaged across shape and height, shadow shading did not have a significant effect on judgments of ground contact in virtual reality.




\paragraph{H3: Is there an interaction between shape and shadow?} 
Analysis of the shadow by shape simple slopes indicated that the effect of shadow was only significant for the sphere ($OR = 1.43$, $p < .05$). The predicted probabilities of correct response for each shape and shadow are displayed in Figure \ref{fig:exp1_slopes} (right).
On average,  when presented with the sphere, people were 76\%~($SE=1.8\%$) accurate with the dark shadow when presented and they were 81\%~($SE=1.6\%$) accurate with the light shadow.

\subsubsection{How difficult was each condition?}
 
We examined post-experiment survey responses to better understand how participants interpreted our experimental stimuli.  Participants were asked to rank the devices from easiest to hardest for determining ground contact, and they were asked to rank devices on graphics quality from lowest to highest. For difficulty rankings, they assigned each XR display to one of three categories: easiest, middle, or hardest. For quality of graphics rankings, they assigned each display to best, middle, or worst. 

There was no clear easiest display, given that all three XR devices were rated as the easiest display by two out of six (2/6) participants. However, the AR devices were seen as more difficult than the VR condition, with both the OST AR display (4/6) and the VST AR display (2/6) receiving ratings for being the most difficult display. Participants also stated that the video see-through display had the lowest perceived quality of graphics with three out of six (3/6) worst votes and 3/6 median votes,  and that the optical see-through display had the highest quality graphics (5/6). 


Peoples' reported strategies in the post-experiment survey support the idea that the curved lines of the sphere and its shadow increased the difficulty of discerning surface contact. For discussion, we will refer to each participant with an acronym (e.g., P1 for the first participant).  P1, P3, and P4 reported looking at shadows ``\textit{beneath}" the objects or ``\textit{beneath the front side}" of objects. P3 also commented that they tried to remember the point of optical contact for a given object. P2 and P5 explicitly stated that they looked for the ``\textit{edges}" of shadows to make judgments. A reliance on shadow edges would also explain why people's responses for the sphere were generally less accurate, regardless of XR display. Participants reported that they employed the same strategies for discerning ground contact across devices.

\subsection{Discussion}

In our first experiment, we found that light shadows improved ground contact judgments compared to dark shadows for both AR devices and that they performed comparably to dark shadows in VR. These results encourage the use of non-photorealistic rendering solutions for improving surface contact judgments across XR displays. Perceived surface contact plays an important role in depth perception and improving it may help alleviate the disconnected appearance between virtual objects and real surfaces in AR.

In addition, we found that object shape influences surface contact judgments, where judgments were more accurate for the cube than the sphere in all three XR devices. Judgments for the icosahedron were more accurate than the sphere in VST AR and OST AR. Given that there was only one condition where people's judgments were more accurate for the sphere than another target shape, our current study provides some evidence that spheres may be less effective than other shapes in establishing surface contact. This finding may have implications for depth perception in XR studies, as spheres are a common target object used in distance perception research for these devices \cite{Armbruster:2008:DPV, Gao:2019:IVO, Peillard:2019:SED, Peillard:2020:CRP}. In addition, while our results suggest an overall advantage for light shadows for AR, there were some differences in these effects for specific shapes and devices. The benefit of light shadows was largely driven by the cube in OST AR. But, in VST AR all shapes were significantly influenced by the presence of a shadow. 



There are limitations to the present experiment. All objects were viewed from a single vantage point, which may affect how generalizable our results are to other geometric shapes and perspectives in XR. This concern is supported by prior research, which has demonstrated that spatial perception can be influenced by viewing angle \cite{Lawson:2008:UMF,Ahn:2019:SPA}. Some prior behavioral studies in AR have made conscious efforts to display objects at multiple orientations to subvert any possible effects of orientation \cite{Gao:2019:IVO}. Others have displayed objects at varying heights and distances \cite{Adams:2021:SLC}. Thus, to understand better how the placement of an object may affect people's ability to determine surface contact, we will evaluate target objects at different orientations in the next experiment. 

A second limitation is that we do not evaluate any 3D objects that have a comparable complexity to those that might be used in typical AR and VR applications. All three objects that we evaluated were simple geometric shapes. As highlighted by Powell et al. \cite{Powell:2012:VPV}, an object's geometry has the ability to influence one's perception of space. In Powell and colleagues' study, researchers evaluated people's ability to reach and grasp virtual spheres, icosahedrons, and apples in VR. They found that the richer geometry cues provided by the icosahedron and apple positively influenced reaching and grasping behaviors. In a similar manner, Do et al. \cite{Do:2020:EOS} evaluated the effect of object color and luminance on objects of different shapes for depth judgments in mobile AR. They found interactions with color and luminance on depth perception. The findings of both of these studies motivate us to evaluate a more complex geometric shape in our next experiment.

\begin{figure}[b]
    \centering
    \includegraphics[width=\linewidth, keepaspectratio]{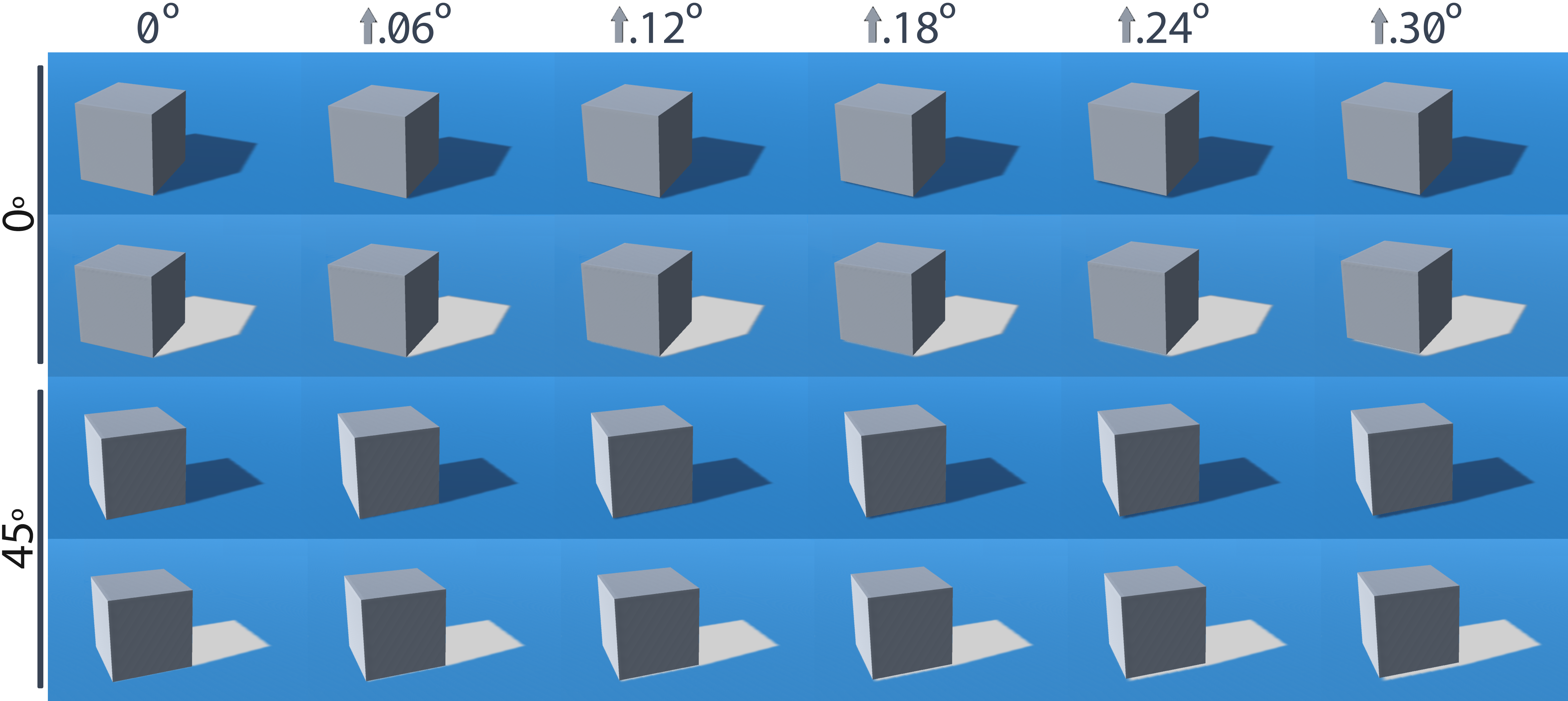}
        \caption{The cube from Experiment 1 is rotated by 45$^{\circ}$ and evaluated under the same shadow shading methods and vertical displacements in Experiment 2.  Here both cubes are shown for visual comparison.}
       \label{fig:exp2_cubes}
\end{figure}

\section{Experiment 2: Complexity and Orientation} \label{exp2}

\begin{figure}[b]
    \centering
       \includegraphics[width=\linewidth, keepaspectratio]{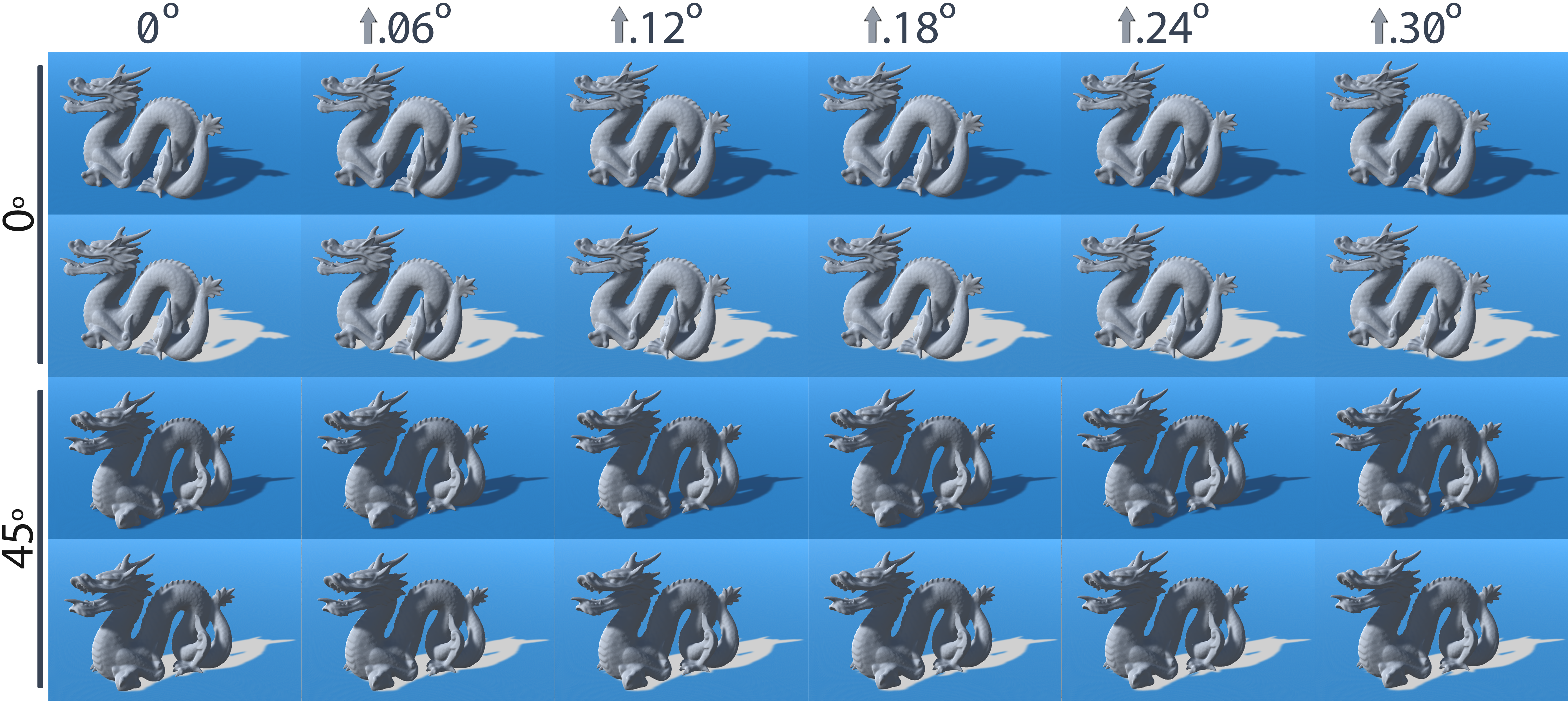}
        \caption{In Experiment 2, the Stanford dragon was also rendered in each display and shown at two orientations. All other shading and displacement factors were the same as in Experiment 1.}
        \label{fig:exp2_dragons}
\end{figure}

In this experiment, we investigated if shape complexity and orientation influenced the perception of surface contact when dark and light shadows were present.  Figures~\ref{fig:exp2_cubes} and~\ref{fig:exp2_dragons}, which display the experimental stimuli used in Experiment 2, present a strong visual argument for investigating orientation given we can observe changes in shape and shading between an object and its cast shadows in these images with rotation. To address these concerns, we evaluated two target objects, a cube and the Stanford dragon, at two orientations, rotated by $0^{\circ}$ and rotated by $45^{\circ}$, in our second study. Our hypotheses for these experiments were as follows:

\textbf{H4}: 
In our first experiment, people's surface contact judgements were more accurate for the cube than the sphere across all XR devices. Although the sphere was technically a more complex geometric shape \cite{Fox:1998:PAS,Do:2020:EOS}---where geometric complexity is defined by the number of polygons used to construct a given object's mesh---the resulting shape was nonetheless a geometric primitive. As such, in our second experiment we evaluated a notably more complex object, the Stanford dragon, which consisted of 113,000 polygons. 

Although the dragon object provides additional depth from shading information due to its complexity, we anticipated that people's responses to the cube would be more accurate, given the importance of contour junctions (e.g., T-junctions) as local cues for occlusion \cite{Rubin:2001:RJS} when perceiving surfaces \cite{Nakayama:1992:EPV}. Accordingly, the straight, rectilinear edges of the cube should prove more beneficial to ground contact judgments.
This finding would provide evidence against the idea that more complex geometries are inherently better for depth perception \cite{Do:2020:EOS,Powell:2012:VPV,Bailey:2006:REW}. 

\textbf{H5}: Based on the results from our first study, we once again predicted that people's surface contact judgments would be more accurate for the light shadow condition than the dark shadow condition across both AR devices. 

\textbf{H6}: We predicted that, overall, a change in object orientation would affect participants' surface contact judgments given that changes in orientation can alter both the perceived shape of the surface contact and its associated cast shadow for the viewer. Because the starting position ($0^{\circ}$) was different for the cube and the dragon, it's possible that the $45^{\circ}$ rotation might have deferentially affected the perception of the shapes. Therefore, we also include an interaction between shape and orientation in our analysis.




\subsection{Participants}
The same six individuals (3M, 3F) aged 23--30 volunteered to participate in our second experiment. Our experimental methods were approved by the local institutional review board, written consent was obtained from all subjects prior to participation, and each participant was paid 20 USD for 2-3 hours of their time.

\begin{figure*}[hb!]
    \centering
        \includegraphics[width=0.325\linewidth, keepaspectratio]{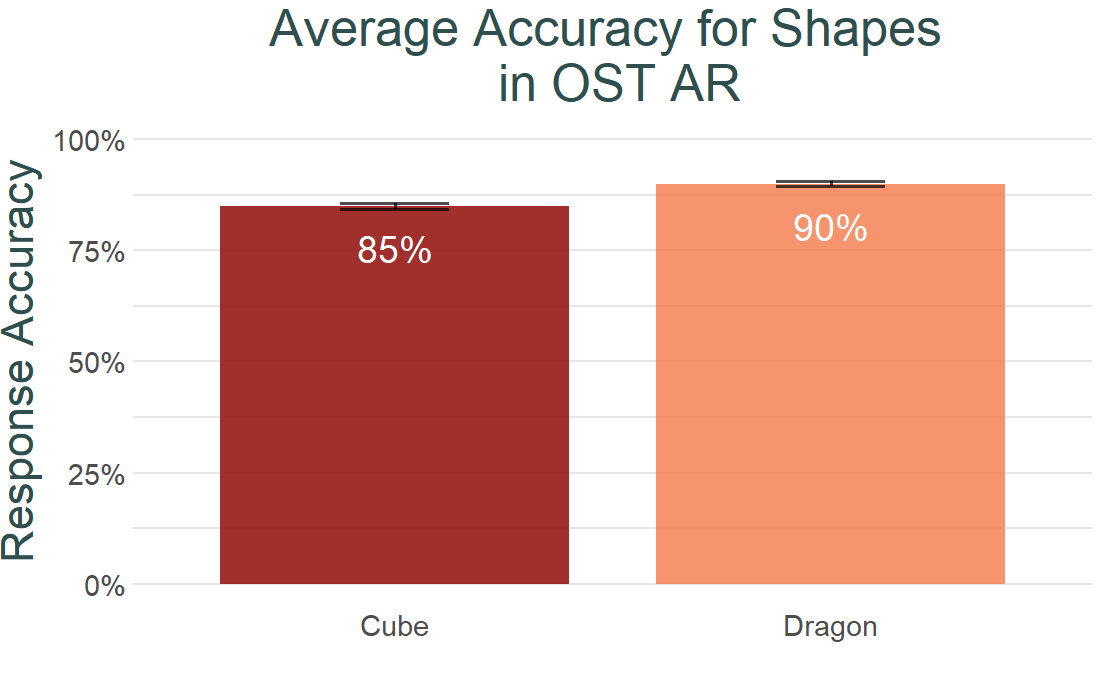}
        \includegraphics[width=0.325\linewidth, keepaspectratio]{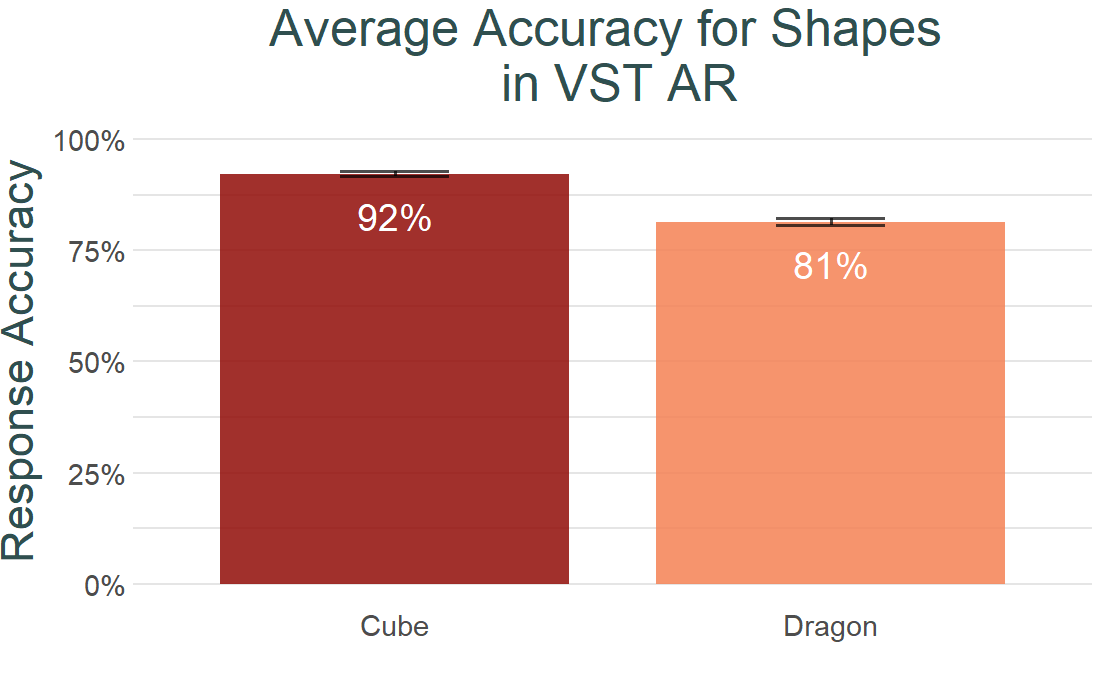}
        \includegraphics[width=0.325\linewidth, keepaspectratio]{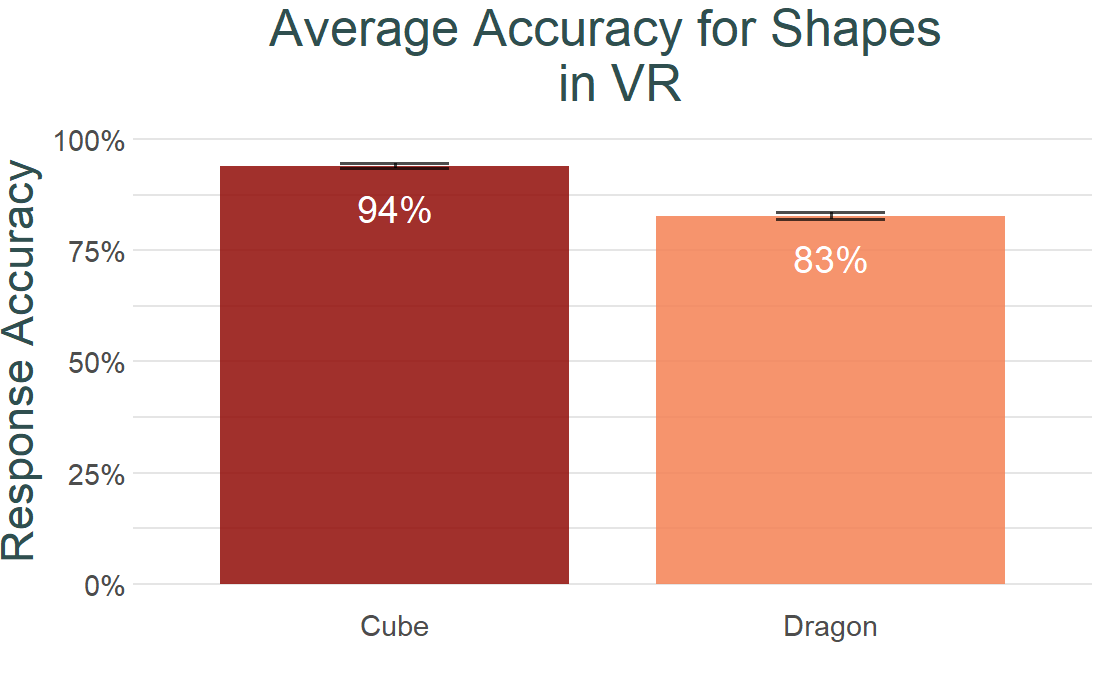}
        \caption{Experiment 2 --- Average percentage of correct responses for each target object shape by display condition: optical see-through AR (left), video see-through AR (center), and virtual reality (right). Object complexity did not inherently benefit surface contact judgements. People's judgements were significantly more accurate when presented with the cube than the dragon in both the VST AR and VR conditions. The opposite was true in OST AR. }
        \label{fig:exp2_shape_results}
\end{figure*}

\subsection{Design}

Our experimental design for the second evaluation mirrored the one employed by our first, in which a temporal two-alternative forced choice (2AFC) paradigm with a method of constant stimuli was employed. All experimental details were the same, except for the independent variables. We used the same testing environment, height displacement values (between 0 and 6mm), the same number of comparisons, and the same number of trials. We employed two models, the cube from Experiment~1, and the Stanford Dragon,\footnote{http://graphics.stanford.edu/data/3Dscanrep/} tesselated to have 113,000 polygons. 

Each participant evaluated two cast shadow conditions (dark and light) over three different object + orientation conditions: the cube rotated by $45^{\circ}$, the Stanford dragon rotated by $0^{\circ}$, and the Stanford dragon rotated by $45^{\circ}$. This resulted in 6 unique combinations of experimental stimuli. However, our analysis evaluated 8 combinations of stimuli, as we included the non-rotated cube condition (with both shadow conditions) from Experiment 1. The Experiment 1 data was included to prevent participant fatigue from an increased number of trials and because the participants were the same, so concerns about individual differences overall are low. In the second experiment, 1,980 points of data were collected from each volunteer for a total of 11,880 data points. When combined with the cube data from the first experiment, each participant completed 2,640 trials for a total of 15,840 trials across all participants. 


Using the same approach that we employed for Experiment 1, we evaluated the extent to which participant variance accounted for variance in our collected data by calculating the intraclass correlation coefficient (ICC) for Experiment 2. For the OST AR display condition we found $\tau_{ost}$ = 0.006 and for the VST AR display we found $\tau_{vst}$ = 0.028. For the VR display condition $\tau_{vr}$ = 0.023. Because our ICC values were near zero, we determined that our experimental findings were not significantly influenced by between-participant variation.

\subsection{Procedure}
The procedure for the second experiment followed the same protocol used in Experiment 1. However, this time volunteers did not undergo training for the devices and experimental paradigm since they had already performed a very similar experimental task in Experiment 1.


\begin{table}[t]
\caption{
Experiment 2 --- The accuracy for each shape and shadow condition tested reported with the standard errors. } 
\label{tab:exp2_acc}
\centering
\begin{tabular}{lllcc}
\multicolumn{5}{c}{\textbf{Accuracy for Experiment 2}} \\ 
\toprule
           Condition &           &\multicolumn{1}{c}{OST AR}   &\multicolumn{1}{c}{VST AR}   &\multicolumn{1}{c}{VR}  \\
           \\ 
     Cube &     &85\% $\ [\ .7\%]$  &92\% $\ [\ .5\%]$  &94\% $\ [\ .5\%]$  \\
     Dragon &   &90\% $\ [\ .6\%]$  &81\% $\ [\ .8\%]$  &83\% $\ [\ .8\%]$  \\
\midrule \\   
     Dark &    &81\% $\ [\ .8\%]$  &83\% $\ [\ .8\%]$  &86\% $\ [\ .7\%]$  \\ 
     Light &   &94\% $\ [\ .5\%]$  &91\% $\ [\ .6\%]$  &90\% $\ [\ .6\%]$  \\
\midrule \\   
     Origin ($0^{\circ}$)   & &84\% $\ [\ .7\%]$  &87\% $\ [\ .7\%]$  &86\% $\ [\ .7\%]$  \\
     Rotated ($45^{\circ}$) & &91\% $\ [\ .6\%]$  &87\% $\ [\ .7\%]$  &91\% $\ [\ .6\%]$ \\
\bottomrule
\end{tabular}
\end{table}

\subsection{Results} \label{exp2_results}



\paragraph{Statistical Analyses} 

For our analyses, we used the binary logistic regression model approach that we employed in the first study (See Section \ref{exp1_results}) to analyze participants' 2AFC judgements for each of the three XR devices.  Separate GLMs with logit link functions were run for each device to understand how an object's cast shadow shading, geometric complexity, and orientation affected people's surface contact judgments in each device. Specifically, we modeled binary outcomes (correct or incorrect) for our predictors: object shape, object orientation, shadow shading, and height. Height was recorded in millimeters then centered at zero and treated as continuous. Object shape (2 levels: cube and dragon), orientation (2 levels: rotated by $0^{\circ}$ and by $45^{\circ}$), and shadow shading (2 levels:  dark and light) were treated as categorical factors. These factors were deviation coded. For shadow shading, dark was coded as -.5 and light was .5; for shape, cube was -.5 and dragon was .5; for orientation, $0^{\circ}$ was -.5 and $45^{\circ}$ was .5. 

We tested three planned comparisons for each device. These comparisons were:
 1) whether surface contact judgments differed for different object geometries (i.e., \textbf{H4} the main effect of object shape); 2) the difference in surface contact judgments between dark and light shadow shading across all other conditions (i.e., \textbf{H5} the main effect of shadow shading); and 3) whether an object's orientation influenced people's surface contact judgments (i.e., \textbf{H6} the main effect of orientation). We also included an interaction between object shape and shading and an interaction between object shape and orientation to better understand the relationship between these variables.  



\begin{table*}[t]
\caption{Experiment 2 --- Results of planned comparisons using binary logistic regression models for each display condition are displayed. \emph{B} is the regression coefficient, $SE_{B}$ is the standard error of the regression coefficient, $OR$ is the odds ratio, and $CI_{OR}$ is the confidence interval associated with the odds ratio. Negative values for \emph{B} indicate that the first factor in the comparison was more accurate, whereas positive values indicate that the second factor was more accurate. 
} 
\label{tab:exp2_results}
\resizebox{\linewidth}{!}{%
\begin{tabular}{ll SlSl SlSl SlSl}
\\
\multicolumn{14}{c}{\textbf{Results for Experiment 2}} \\
\toprule  
\\
&& \multicolumn{4}{c}{\hspace{22pt}{OST AR}}  &\multicolumn{4}{c}{\hspace{22pt}{VST AR}}  &\multicolumn{4}{c}{\hspace{22pt}{VR}} 
\\ \\
    Predictor  &
    &\emph{B} &$SE_{B}$  &$OR$ &\multicolumn{1}{c}{$[CI]_{OR}$}
    &\emph{B} &$SE_{B}$  &$OR$ &\multicolumn{1}{c}{$[CI]_{OR}$} 
    &\emph{B} &$SE_{B}$  &$OR$ &\multicolumn{1}{c}{$[CI]_{OR}$} \\
\cmidrule(lr{12pt}){1-2}  \cmidrule(l{22pt}r){3-6} \cmidrule(l{22pt}r){7-10} \cmidrule(l{22pt}r){11-14} \\ 
    Shape 
    &(cube vs dragon)                     
        &.32$\ $**      &.11   & 1.38   & $[1.12, \ 1.70]$   
        &-1.15$\ $***   &.10   & .32    & $[\ .26,\quad  .39]$ 
        &-1.31$\ $***   &.11   & .27   & $[\ \ .22,\ \ \ .33]$  \\
    Shadow 
    &(dark vs light)    
        & 1.31$\ $***   &.10  & 3.70   & $[3.02,\ 4.55]$  
        & .91$\ $***   &.10  & 2.47   & $[2.03,\ 3.04]$ 
        & .52$\ $***   &.11  & 1.68   &$[1.36,\ 2.08]$   \\ 
    Orien 
    &($0^{\circ}$ vs $45^{\circ}$)
        &.67$\ $***    &.10   & 1.96  & $[1.62, \ 2.38]$   
        &-.06      &.10   & .94   & $[\ \ .78, \ 1.14]$   
        &-.13      &.11   & .88   & $[\ \ .71, \ 1.08]$   
    \\ \\ 
    Shape$\times$Shadow
    &&-.90$\ $***   &.21  &.41  &$[\ \ .27,\ \ \ .61]$
    &-.50$\ $*   &.21  &.61  &$[\ \ .40,\ \ \ .91]$
    &-0.45$\ $*   &.21  &.64 &$[\ \ .42,\ \ \ .97]$ \\
    &(cube:\hspace{7px}  dark vs light)    
        & 1.76$\ $***       &.15     & 5.79    & $[4.35,\ 7.71]$  
        & 1.15$\ $***  &.17    &3.17    &$[2.25,\ 4.46]$
        & .74$\ $***  &.18     & 2.11   &$[1.47,\ 3.02]$ \\
    &(dragon: dark vs light)
    & .87$\ $***      &.15   & 2.37    & $[1.77,\ 3.17]$
        & .66$\ $***  &.11   & 1.93   & $[1.55,\ 2.40]$
        &.29$\ $**    &.11   & 1.34   & $[1.08,\ 1.67]$    
    \\ \\ 
    Shape$\times$Orien
    &&-0.51$\ $**   &.20  &.61  &$[\ \ .41,\ \ .88]$
    &.20     &.19    &1.22    & $[\ \ .84,\ 1.79]$
    &.77$\ $***  &.21    &2.16       &$[1.43,\ 3.29]$ \\
    &(cube:\hspace{7px} $0^\circ$ vs $45^\circ$)  
        &.93$\ $***     &.13    &2.53    &   $[1.95,\ 3.27]$
        &-.16     &.16   & .85     &$[\ \ .63,\ 1.16]$ 
        &-.52$\ $**    &.18   & .60   &$[\ \ .42,\ \ \ .85]$ \\
    &(dragon: $0^\circ$ vs $45^\circ$)                   
        &.42$\ $**  & .14   & 1.52   & $[1.15,\ 2.02]$
        &.04        & .11   & 1.04   & $[\ \ .84,\ 1.29]$ 
        &.26$\ $*   & .11  & 1.29   &$[1.04,\ 1.61]$ 
    \\ \\
    Height&                                  
        &.12$\ $***  &.01   & 1.13   &$[\ \ 1.11,\ \ 1.14]$
        &.09$\ $***  &.01  & 1.09   &$[1.08,\ \ 1.11]$
        &.08$\ $***  &.01  & 1.08   &$[\ 1.07,\ \ 1.10]$ \\
    Intercept&                               
        & 2.50$\ $*** &.07  &12.13  &$[10.71,13.83]$
        &2.25$\ $***  &.06  &9.50  &$[8.51,10.67]$ 
        &2.39$\ $*** &.06  &10.88  &$[\ 9.69,12.28]$  \\ 
\bottomrule 
    &\multicolumn{9}{c}{ }   &\multicolumn{4}{c}{$* p < .5$ \qquad $** p < .01$ \qquad $*** p < .001$} \\
\end{tabular}}%
\end{table*}

\begin{figure*}[b]
    \centering
        \includegraphics[width=0.325\linewidth, keepaspectratio]{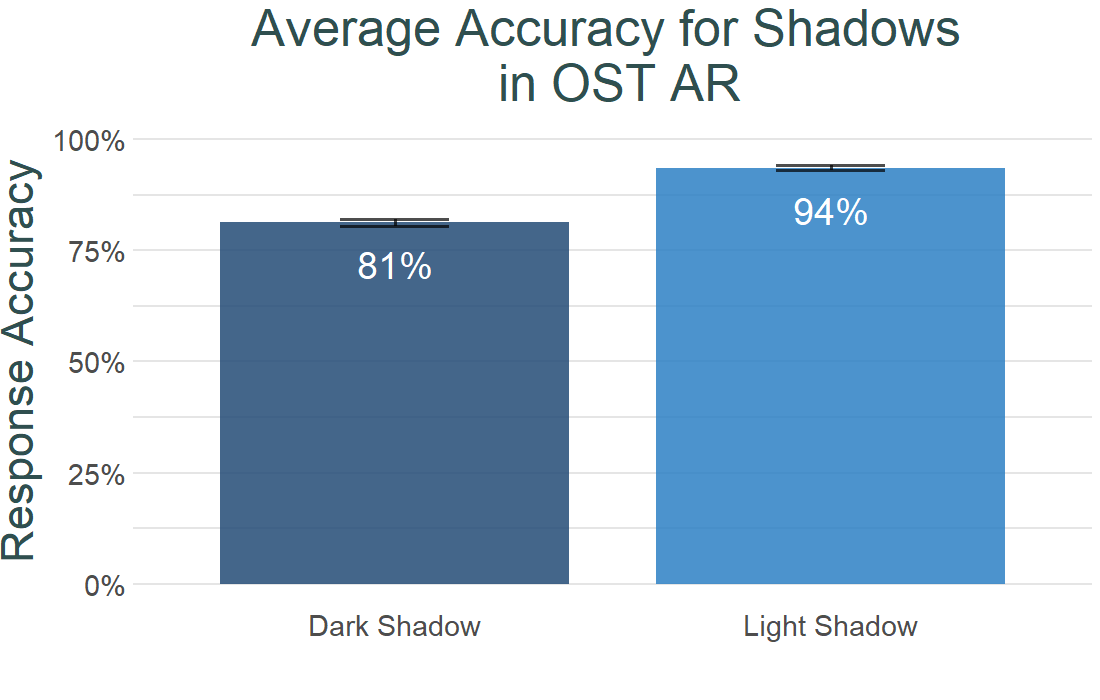}
        \includegraphics[width=0.325\linewidth, keepaspectratio]{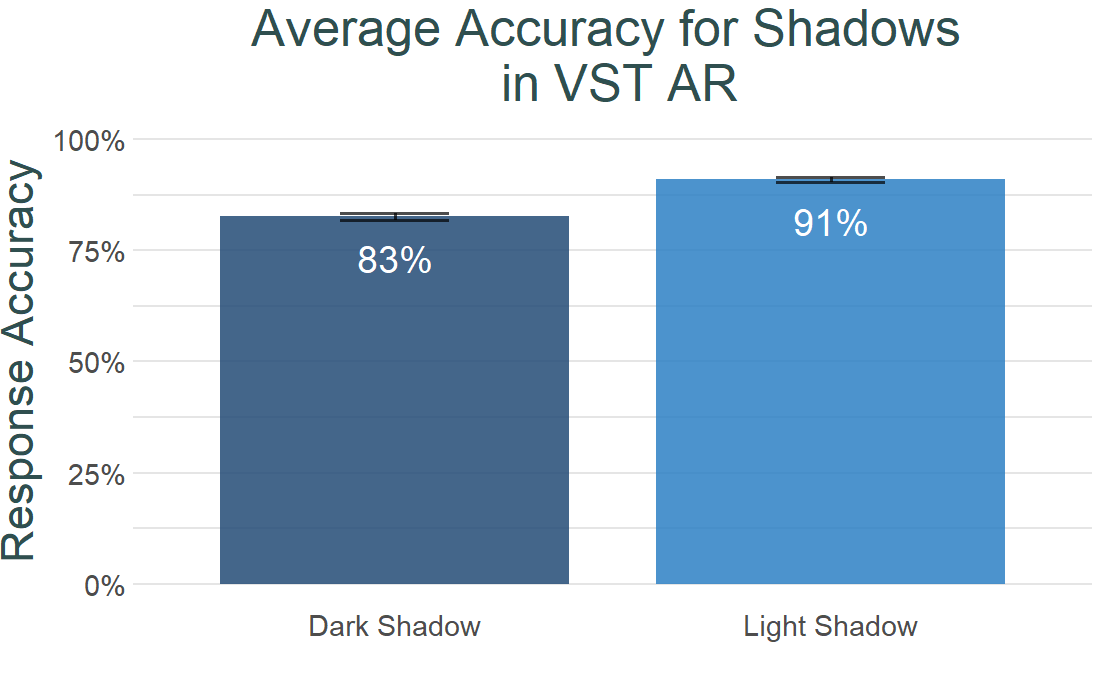}
        \includegraphics[width=0.325\linewidth, keepaspectratio]{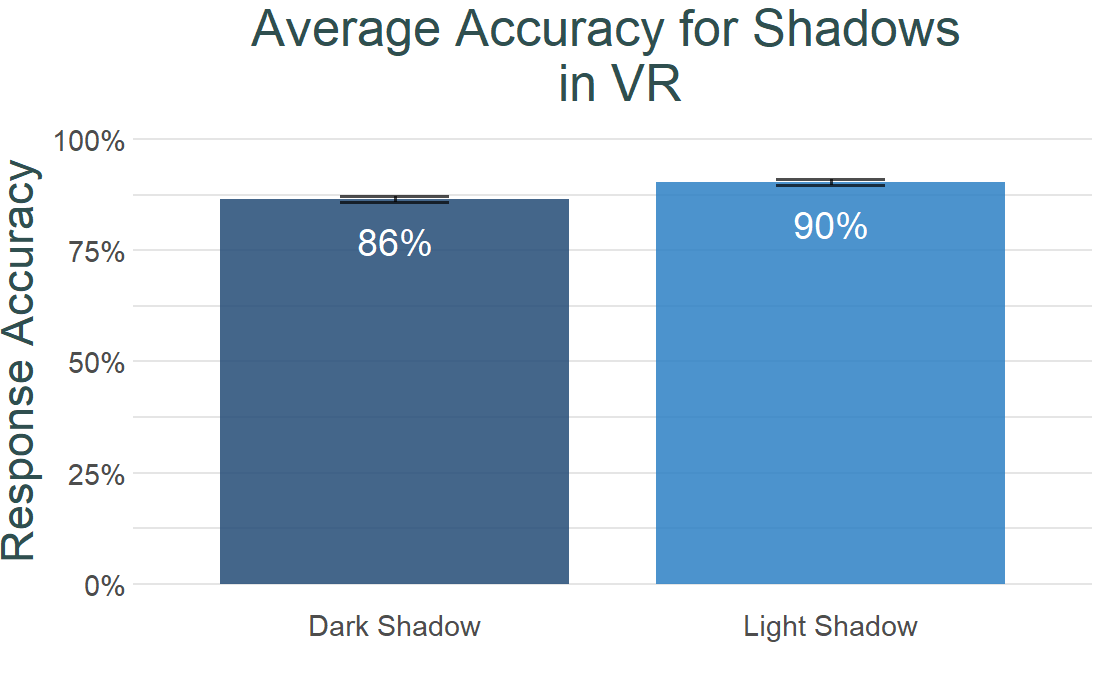}
        \caption{Experiment 2 --- Average percentage of correct responses for each target object shadow by display condition: optical see-through AR (left), video see-through AR (center), and virtual reality (right). People were significantly more accurate when shadow shading was light, rather than dark in all three display conditions. }
        \label{fig:exp2_shadows_results}
\end{figure*}

\subsubsection{Optical See-through AR} \label{exp2:res:ost}
We show the average percent of correct response for each of the evaluated main effects in Table \ref{tab:exp2_acc}, and the results of the logistic regression for the OST AR display data are presented in Table \ref{tab:exp2_results}.  We expected participants to be more likely to correctly indicate which object was on the ground as the height of target objects increased.
The improvement of participants' performance as height increased is demonstrated statistically by a main effect of height in our logistic regression model ($OR = 1.13$, $p < .001$). An odds ratio ($OR$) of 1.13 indicates that for every 1mm increase in height, the odds of correctly stating which object was closer to the ground increased by a factor of~1.13.


\paragraph{H4: Does an object's complexity matter?} 

People were more accurate when assessing surface contact with the dragon than the cube, with 90\% and 85\% accuracy, on average, overall  (Figure~\ref{fig:exp2_shape_results}). 
Our statistical analysis revealed that this relationship was significant given that---when averaged across shadow shading, orientation, and height---the dragon shape in our model was more likely to elicit correct responses than the cube with an odds ratio of 1.38 ($p < .01$). As such, people were 1.38 times more likely to correctly assess surface contact when presented with the dragon. 
This outcome differs from what we observed in both the VST AR and VR conditions as reported in Sections \ref{exp2:res:vst} and \ref{exp2:res:vr}. In these devices surface contact judgments to the cube were more accurate than judgments to the dragon.

\begin{figure*}[ht]
    \centering
    \includegraphics[width=\linewidth, keepaspectratio]{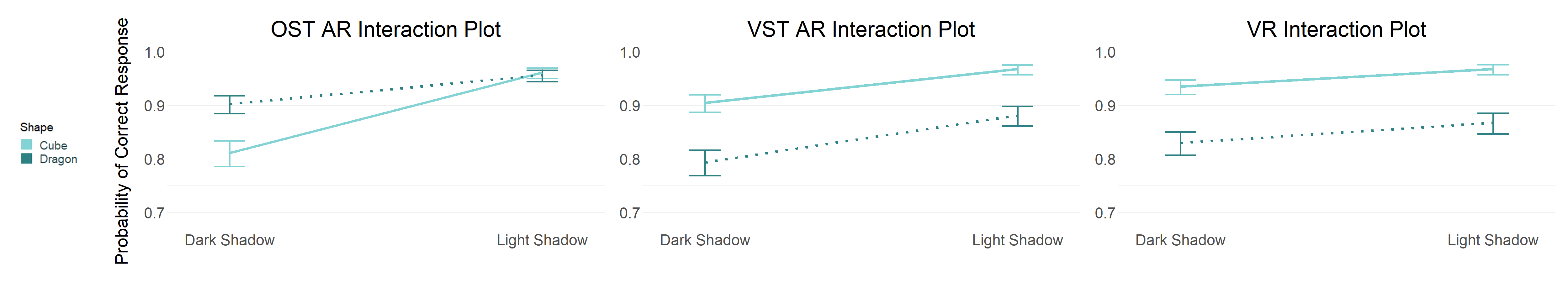}
        \caption{
        Experiment 2 --- Predicted probability of correct response for shape (cube, dragon) and shadow (dark, light) interactions  with 95\% confidence. For all three devices a light shadow was more likely to elicit a correct judgement of surface contact than a dark shadow, regardless of shape. In addition, this effect was stronger for the cube than for the dragon. }
       \label{fig:exp2_slopes}
\end{figure*}

\paragraph{H5: Does shadow shading method matter?} 
The main effect for shadow ($OR = 3.70$, $p < .001$) indicates that a correct response was 3.70 times more likely when the object was presented with a light shadow compared to a dark shadow, when collapsed across shape, orientation, and height. On average, people were 94\% accurate when presented with the light shadow and 81\% accurate when presented with the dark shadow  (See Figure~\ref{fig:exp2_shadows_results}).  
There was also a significant interaction between shape and shadow ($OR = .41$, $p < .001$). Analysis of the shadow simple slopes by shape indicated that, for both shapes, a light shadow was more likely to yield a correct judgment of ground contact than a dark shadow (supporting the main effect of shadow). The simple slopes are depicted in Figure~\ref{fig:exp2_slopes}. However, this effect was stronger for the cube ($OR = 5.79$, $p < .001$) than the dragon ($OR = 2.37$, $p < .001$).

\paragraph{H6: Does object orientation matter?}
There was a significant main effect of orientation ($OR = 1.96$, $p < .001$) indicating that, overall, participants were more likely to make a correct judgment when the object was rotated $45^\circ$ from its starting position. On average, people's judgments to targets at their original orientation were 84\% accurate and their judgments to rotated targets were 91\% accurate in the HoloLens. 
This main effect was qualified by an interaction between object shape and orientation ($OR = .60$, $p < .01$).
In order to interpret the interaction, we calculated the orientation simple slopes for each shape. For both shapes, the likelihood of correctly judging ground contact was higher for the $45^\circ$ orientation than the $0^\circ$ orientation (supporting the main effect of orientation). However, this effect was stronger for the cube ($OR = 2.53$, $p < .001$) than for the dragon ($OR = 1.52$, $p < .01$).




\subsubsection{Video See-through AR} \label{exp2:res:vst}

We show the average percent correct response for each of the evaluated main effects in Table \ref{tab:exp2_acc}.
The results of our statistical analyses are reported in Table \ref{tab:exp2_results}.
For VST AR we again found a main effect of height ($OR = 1.09$, $p < .001$) such that for every 1mm increase in height, the odds of correctly stating which object was closer to the ground increased by a factor of 1.09.  Accordingly, as the height of the vertically displaced object increased, participants were more likely to correctly indicate which object was on the ground.

\paragraph{H4: Does an object's complexity matter?} 
The cube shape was more likely to elicit correct responses than the dragon  ($OR = .32$, $p < .001$), suggesting that object complexity is not inherently beneficial to surface contact perception.  People's judgments were 92\% accurate, on average, when assessing surface contact with the cube and 81\% accurate, on average, when assessing contact with the dragon (Figure~\ref{fig:exp2_shape_results}). 

\paragraph{H5: Does shadow shading method matter?} 
The main effect for shadow ($OR = 2.47$, $p < .001$) indicates that a correct response was 2.47 times more likely when the object was presented with a light shadow compared to a dark shadow. On average, people were 91\% accurate with the light shadow and they were 83\% accurate with the dark shadow when evaluating surface contact (See Figure~\ref{fig:exp2_shadows_results}). 
In addition, there was a significant interaction between shape and shadow ($OR = .61$, $p < .05$). Analysis of the shadow by shape simple slopes indicated that the effect of shadow was significant for both shapes (See Figure~\ref{fig:exp2_slopes} (center)). For both shapes, the probability of correct response was higher when a light shadow was presented compared to a dark shadow. However, this effect was stronger for the cube ($OR = 3.17$, $p < .001$) than the dragon ($OR = 1.93$, $p < .001$). 


\paragraph{H6: Does object orientation matter?}

There was no main effect of orientation in VST AR. People's surface contact judgments to targets when positioned at their original orientations and when rotated were 87\% accurate, on average.  The interaction between orientation and shape was also insignificant. In other words, the probability of providing a correct response was not significantly different between the two orientations in VST AR. This outcome differs from the results of our analyses in the OST AR and VR conditions, both of which found that people's judgments were more accurate on average to the rotated targets. 



\subsubsection{Virtual Reality} \label{exp2:res:vr}

The average rate of correct response for each main condition is reported in Table \ref{tab:exp2_acc}, and the results of our statistical analyses are reported in Table~\ref{tab:exp2_results}.
For VR, we found a main effect of height ($OR = 1.06$, $p < .001$) such that for every 1mm increase in height, the odds of correctly stating which object was closer to the ground increased by a factor of~1.06.

\paragraph{H5: Does an object's complexity matter?} 
The cube was more likely to elicit correct responses than the dragon  ($OR = .27$, $p < .001$), suggesting that object complexity is not inherently beneficial to surface contact perception. Participants' judgments were 94\% accurate, on average, when assessing surface contact with the cube and 83\% accurate, on average, when assessing contact with the dragon (Figure~\ref{fig:exp2_shape_results}). 

\paragraph{H4: Does shadow shading method matter?} 
The main effect for shadow ($OR = 1.68$, $p < .001$) indicates that a correct response was 1.68 times more likely when the object was presented with a light shadow compared to a dark shadow. Judgments of surface contact were 90\% accurate, on average, in VR with the light shadow and they were 86\% accurate, on average, with the dark shadow (See Figure~\ref{fig:exp2_shadows_results}). 
The interaction between shape and shadow was also significant ($OR = .64$, $p < .05$). Unlike our results for Experiment 1, the analysis of the shadow by shape simple slopes indicated that the effect of shadow was significant for both shapes in VR (See Figure \ref{fig:exp2_slopes} (right)). The probability of correct response was higher when a light shadow was presented regardless of the object presented, although the shadow effect was stronger for the cube ($OR = 2.11$, $p < .001$) than the dragon ($OR = 1.34$, $p < .001$).

\paragraph{H6: Does object orientation matter?}
The main effect of orientation was not significant. However, there was a significant interaction between shape and orientation ($OR = 2.16$, $p < .001$). On average, people's judgments to targets at their original orientation were 86\% accurate and their judgments to rotated targets were 91\% accurate.  
Analysis of the orientation by shape simple slopes indicated that a correct response was more likely when the cube was presented at 0$^\circ$ than when the cube was rotated 45$^\circ$ ($OR = .60$, $p < .01$). In contrast, the probability of a correct response was higher when the dragon was rotated 45$^\circ$ than when it was presented at 0$^\circ$ ($OR = 1.29$, $p < .05$).

\subsubsection{How difficult was each condition?}

After completing the second experiment, volunteers were asked to rank each device by difficulty and by quality of graphics as in Experiment 1. All participants ranked the video see-through display as the most difficult device (6/6) and the device with the lowest quality of graphics (6/6). In contrast, the optical see-through display was most consistently rated as the easiest display (4/6) and the display with the highest quality of graphics (5/6). Ratings were also corroborated by comments at the end of the survey. Both P2 and P4 stated that the HoloLens ``\textit{felt more clear}," and P1 commented that ``\textit{the resolution of the Zed was really painful for my eyes."} 

Participants were also asked about the strategies they used to discern surface contact. P3 commented that the dragon was ``\textit{easier but more mentally exhausting... since the dragon moved.}" Their comment implied that they had to visually search for a shadow before making a judgment. Their complaint is reasonable since, as the shape of the dragon was asymmetrical, changes in orientation were more pronounced (See Figure \ref{fig:exp2_dragons}). P3 further bemoaned of the dragon that ``\textit{when too detailed it was overwhelming, [but] when not detailed enough it became equally difficult to use}." P2 commented that the dragon provided more ``\textit{pockets of shadow,"} which altered their strategy such that ``\textit{ It wasn't just watching an edge, more like watching for shape and quantity of shadow.}" This comment hints that people may have been able to use additional depth information provided by the dragon (e.g., shape from shading) to inform their judgments of surface contact.



\subsection{Discussion} \label{exp2_discussion}

In Experiment 2, for both AR display conditions, we again found that judgments of surface contact benefited from the presence of non-photorealistic, light cast shadows. However, in contrast to our first study, we also found that light shadows improved surface contact judgments in VR. Looking at shape $\times$ shadow interactions for the cube in Experiment 1 and Experiment 2 in VR may provide some insight into this difference in shadow effectiveness for VR between experiments. Both experiments evaluated the same cube object, but the object was presented at a new orientation in Experiment 2. Because we found a significant difference in orientation for the cube and because we found an effect of shadow shading condition in Experiment 2, we infer that the change in orientation for the cube made the light shadows more important in VR. 





Our hypothesis that surface contact judgments for the dragon would be worse than the cube was partially supported. Although we found better performance with the cube compared to the dragon in both VST AR and VR, we found better performance with the dragon in OST AR. With the current experiment alone, we are unable to explain why judgments for the dragon in the OST AR device were significantly more accurate than those for the cube. However, the current study provides evidence that more complex geometries do not inherently benefit surface contact perception. Although complex 3D shapes can provide more depth information via self-shading cues, it appears that other factors such as the bottom edge of the object may be more important. However, additional future research is needed to verify this claim. 



Given the effects of orientation that we found in both the OST AR and VR displays, it will be important to consider the orientation of objects in XR in future application development. This finding is in line with prior research which shows that the angle from which we view a 3D object may affect spatial perception judgments related to that object\cite{Lawson:2008:UMF,Ahn:2019:SPA}.  The lack of difference in orientation for the video see-through conditions requires further investigation.

\section{General Discussion}\label{conclusions}

Perception of surface contact is important for determining the scale of space and for governing our interactions within space. Cast shadows provide important information about surface contact, and thus provide important cues to our visual system for perceiving and acting with our surroundings \cite{Thompson:1998:VG,Ni:2004:PSL,Adams:2021:SLC}. In certain types of XR displays, rendering cast shadows is difficult. Therefore, this paper manipulated cast shadow rendering across different object shapes in XR displays. Our goal was to see if a unified framework would emerge to help designers better design for accurate spatial perception in XR, given the difficulties with rendering realistic, dark shadows in current technology.

We hypothesized that people would be more likely to correctly perceive surface contact when shadows were illuminated with light rather than dark color values. This hypothesis was based on prior work~\cite{Kersten:1997:MCS,Adams:2021:SLC}. We found support for these hypotheses (\textbf{H2} and \textbf{H5}) for two types of AR devices across two experiments. We also hypothesized that surface contact judgments would be complex and affected by the shape of an object associated with a given surface (\textbf{H1}). This hypothesis was derived from the idea that if a shadow provides visible glue with a surface, then the shape of the object to which that glue is applied should interact with shadow effectiveness \cite{Thompson:1998:VG,Madison:2001:UIS}. We found support for this hypothesis as well, although we evaluated relatively simple object shapes in our first study. 

In our second study, we directly investigated the effect of an object's complexity and orientation on surface contact judgments. Consistent with the reasoning above, we hypothesized that orientation would affect the perception of ground contact (\textbf{H6}). We found that orientation mattered for surface contact judgments in OST AR and VR, but we did not find a difference in VST AR, only partially confirming this hypothesis. Our results suggest that future investigations of surface contact and depth perception may be better able to generalize their findings if targets are presented at multiple orientations like in Gao et al. \cite{Gao:2019:IVO}. 

Given prior research on the importance of contour junctions when perceiving surfaces, we also hypothesized that a more complex geometric shape would not inherently benefit surface contact judgements~(\textbf{H4}). Instead, we predicted that people's contact judgements for a cube, which provides clear T-junctions between itself and the surface beneath it, would be more accurate than judgements for a shape with more complex contour junctions.
Our hypothesis was confirmed in VST AR and VR, but not for OST AR, where the opposite was true. This outcome supports the idea that more complex shapes do not inherently benefit surface contact perception, but it is less informative about what properties of the two evaluated objects led to different outcomes for different displays.

One general finding that we can extract from this set of experiments is that the realistic shadows are not requisite for accurate surface contact perception. Further, non-photorealistic (light) shadows will likely work well across XR devices. Our findings give XR designers some latitude for general design, as they provide evidence that more stylized graphics may used in XR applications without detracting from people's spatial perception. 
In addition, there are many AR applications in which accurate spatial perception may be more important than graphical realism (e.g., AR games \cite{Haller:2004:NPR, Keo:2017:GSV} or training applications \cite{Ho:2020:RWV,LopezMoreno:2010:SDI}). 

Our findings on shape, complexity, and orientation are more complicated. It may be that they do not generalize well across XR devices and that general guidelines are not appropriate here. Clearly, more work is needed on shape and orientation, but our results do offer some initial suggestions. For object shape, we found only one case across all three XR devices where judgments were more accurate for the sphere than another target shape (e.g., sphere compared to the icosahedron in VR). Do et al. \cite{Do:2020:EOS} found that depth judgments to spheres were more sensitive to changes in luminance than other shapes in mobile AR. Both our results and those of Do et al. \cite{Do:2020:EOS} may caution against the use of spheres for AR applications that require accurate spatial perception. Finally, our findings on orientation may have implications for the development of XR applications. The angle from which we view a 3D object can affect spatial perception judgments related to that object \cite{Lawson:2008:UMF,Ahn:2019:SPA}.  For the development of XR applications, however, understanding how specific shapes and orientations of objects influence where people perceive them to be positioned in space may be important.

\subsection{Limitations}



A limitation of the current work is that we did not manipulate object shading in our experiments. Prior research on surface contact perception in XR has evaluated the effect of differences in object and shadow shading on people's surface contact judgments \cite{Adams:2021:SLC}.  Based on this prior research, we elected to use a median gray value to shade all of our test objects to mitigate any effects of target object shading, although we recognize that this reduces generalizability. Further, neither the current work nor the prior work conducted by Adams et al. \cite{Adams:2021:SLC} manipulated the color of background surfaces on which target objects were displayed. Yet just as an object's shading color may influence our spatial perception of that object in space \cite{Do:2020:EOS, Diaz:2017:DDP,Ping:2020:ESM,Ahn:2019:SPA}, so too may the backdrop upon which it is presented \cite{Ozkan:2010:BSH}. In our case, we used the same, median blue tablecloth in both experiments. Further work will need to test background surfaces of varying color and texture to fully understand how these may interact with shadows in XR devices.

It is also possible that the decision to include cube data from Experiment 1 in our analysis for Experiment 2 may have affected our analysis of orientation since cube orientation varied between the two sessions. Specifically, it is possible that training effects could have occurred between sessions since the same cube object was used in both experiments. However, on average people's surface contact ratings for the cube in the second session were less accurate for both the VST AR ( $93\% \rightarrow 92\%$) and VR ($95\% \rightarrow 94\%$) conditions. Accordingly, we believe that participants' responses varied due to changes in orientation---not due to training effects.

\section{Conclusions}


This paper demonstrates that there are advantages to non-photorealistic rendering of cast shadows for surface contact judgments in XR. We found that judgments were better with light shadows in two types of AR devices and in VR under certain conditions. In arriving at this finding we experimented with a variety of object shapes, orientations, and complexities. Our findings suggest that under certain circumstances it may be desirable to use light shadows in XR applications in order to denote surface contact. 

Because the light shadows in our current study enhanced people's accuracy in surface contact judgments---it may be worthwhile to evaluate whether more ambitious non-photorealistic rendering approaches facilitate spatial perception in AR. Evaluating colorful shadows, like those designed by Ooi et al. \cite{Ooi:2020:CCS}, for AR may be a desirable starting point. If more research findings are able to confirm the benefits of non-photorealistic rendering for spatial perception in AR, then we may encourage the use of more stylized graphical elements for designers and developers of XR applications. Intriguing design studies like the one conducted by Sun et al. \cite{Sun:2016:SSD} that enhance users' ability to position objects in 3D space using non-photorealistic surface contact cues are already underway.

However, we must be cautious about arbitrarily applying non-photorealistic effects to XR applications before understanding how they affect perception. Our work and recent research studies conducted by other groups \cite{Sun:2016:SSD,Vaziri:2017:IVE,Adams:2021:SLC} encourage the use of stylized graphics in AR, but these results do not imply that all non-photorealistic rendering techniques will be similarly beneficial. The results of Cidota et al. \cite{Cidota:2016:UVE}, in which blur and fade were shown to have adverse effects on action-based depth judgments in AR within personal space, provides some evidence towards the need for caution. Interestingly, much of the theoretical work on perception in augmented reality has speculated the graphics would need to match reality for accurate depth perception \cite{Drascic:1996:PIA,Kruijff:2010:PIA,Adams:2020:RCC}. But more recent empirical findings provide counter evidence for this idea.

Our current findings may have ramifications for egocentric depth judgments in XR displays given prior distance perception research in which people's egocentric distance estimates to targets were more accurate when cast shadows were present \cite{Ni:2004:PSL}. In future research, it may be worthwhile to evaluate how object shape and cast shadow shading manipulations affect more direct measures of depth perception. Rosales et al. \cite{Rosales:2019:DJO} demonstrated that, in the absence of cast shadows, people perceive an object that is placed above the ground incorrectly as farther away. This may help explain some of the effects of overestimation found in prior AR depth perception research, especially given that many studies use floating objects in their assessments \cite{Peillard:2019:SED,Swan:2015:MRD,Singh:2012:DJR}. In future work, we intend to evaluate how photorealistic and non-photorealistic shadows affect egocentric depth judgments in XR displays.

\acknowledgments{
The authors wish to thank William B. Thompson for his insight and Sam Halimi for his wit. This work was supported by the Office of Naval Research under grant N00014-18-1-2964}

\bibliographystyle{abbrv-doi}

\bibliography{ar.bib,vr.bib,perception.bib,stats.bib,bobby.bib}

\end{document}